\documentclass[letter]{aastex631}
\usepackage[utf8]{inputenc}

\usepackage{graphicx}	
\usepackage{amsmath}	
\usepackage{amssymb}	

\usepackage{hyperref}

\begin{document}

\title{Proton Synchrotron Origin of the Very High Energy Emission of GRB 190114C}

\author[0000-0002-8422-6351]{Hebzibha Isravel}
\affiliation{Ben-Gurion University of the Negev \\
Beer-Sheva 8410501, Israel}

\author[0000-0001-8667-0889]{Asaf Pe'er}
\affiliation{Bar-Ilan University\\
Ramat-Gan 5290002, Israel}

\author[0000-0003-4477-1846]{Damien B\'egu\'e}
\affiliation{Bar-Ilan University\\
Ramat-Gan 5290002, Israel}

\begin{abstract}

We consider here a proton-synchrotron model to explain the MAGIC
observation of GRB 190114C afterglow in the energy band $0.2 - 1$~TeV, 
while the X-ray spectra are explained by electron-synchrotron 
emission. Given the uncertainty of the particle acceleration process, 
we consider several variations of the model, and show that they all 
match the data very well. We find that the values of the uncertain 
model parameters are reasonable: {isotropic} explosion energy $\sim 
10^{54.5}$~erg, ambient density $\sim 10-100~{\rm cm^{-3}}$, and 
fraction of electrons/protons accelerated to a high energy power law 
of a few per-cents. All these values are directly derived from the 
observed TeV and X-ray fluxes. { The model also requires that protons be accelerated to observed energies as high as a few $10^{20}$~eV. Further, assuming that the jet break takes place after $10^6$~s gives the beaming-corrected energy of the burst to be $\approx 10^{53}$~erg, which is one to two orders of magnitude higher than usually inferred, because of the small fraction of electrons accelerated.}  { Our modeling is} consistent with both late time 
data at all bands, from { optical} to X-rays, and with numerical models of 
particle acceleration. Our results thus demonstrate the relevance of 
proton-synchrotron emission to the high energy observations of GRBs 
during their afterglow phase.
\end{abstract}


\section{Introduction} \label{sec:intro}

While gamma-ray bursts (GRBs) are well known to emit in the energy band keV~$ \leq
\epsilon_\gamma\leq$~GeV, in recent years there are accumulating observations at
even higher energies, in the range of several GeV to TeV. Early detections in this
band date back to
the 1990's. The Energetic Gamma-Ray Experiment Telescope (EGRET) on-board CGRO,
was the first instrument to monitor a photon with an energy of
$\sim18\hspace{0.3em}$GeV in coincidence with GRB 910503 \citep{1992A&A...255L..13S, 1993ApJ...413..281B,
1994Natur.372..652H}, followed by a $\sim1\hspace{0.3em}$TeV photon
observation associated with GRB 970417a in the Milagrito field of
view  \citep{2003ApJ...583..824A}. Successive detections of 14 bursts/year,
on average, have been done by \textit{Fermi}-LAT (Large Area Telescope) in
a broad range of energies up to $\sim 100$~GeV \citep{2018IJMPD..2742003N}.
Very high sub-TeV band detections were recently
reported by the ground-based Cherenkov
observatories, namely the High Energy Stereoscopic System  (HESS) and the Major Atmospheric
Gamma Imaging Cherenkov (MAGIC). HESS is actively detecting GRBs
at energies larger than $100$~GeV, e.g., the 100 to 440 GeV photons reported in  GRB
180720B \citep{2019Natur.575..464A}. Similarly, MAGIC is
also exploring the window of VHE emissions with the detection of
GRB 190114C in the sub-TeV band \citep{MAGIC:2019irs}. 

These recent discoveries have  called for much attention
as GRB afterglow spectra were long predicted to have VHE
signatures of this kind \citep{1994ApJ...432..181M, Dermer_2000, 2002ARA&A..40..137M, 2004RvMP...76.1143P, 2015PhR...561....1K}.  Furthermore, such detections of GRB
photons with energy larger than 1 TeV are expected to become ubiquitous in the near
future, once the upcoming new generation high sensitive $\gamma-$detector Cherenkov
Telescope Array (CTA) begins operations with its unprecedented precision \citep{knodlseder2020cherenkov}.

Of particular interest is the recent detection of the very high energy ($0.2-1\hspace{0.3em}$
TeV) emission from GRB 190114C by MAGIC. Thanks to its proximity, this burst turned to be
one of the brightest bursts ever detected. It was observed by diverse space
observatories from near the Fornax constellation 4.5 billion light years away  with
a redshift of $z=0.4245 \pm 0.0005$ \citep{web:nasa:23708, web:nasa:stats} in a dense environment right in the middle of the luminous galaxy. 
It was reported with the prompt phase isotropic equivalent energy $E_{\rm
iso} \approx 2.5 \times 10 ^{53}$~erg 
for a duration of $T_{\rm 90}\sim 25$~s.
The afterglow of this burst was observed in an unprecedentedly large energy band for
an unprecedented duration. The broadband  afterglow spectrum can be split into a \textit{low-energy} component ranging from optical
to X-rays and a \textit{high-energy} component (MeV to GeV) extended
by a non-thermal VHE tail (at least up to the TeV band)
\citep{1997Natur.389..261F, 1997Natur.386..686V, 2000ApJ...543..722L, zhang_2018}.
The low energy spectral component is
usually observed to be a broken power-law
\citep{costa1997discovery, van1997transient, wijers1997shocked, harrison1999optical,
2000ApJ...529..635B}. It can be interpreted in the framework of synchrotron emission
from shock accelerated relativistic electrons which naturally leads to a broken
power-law spectrum \citep{1993ApJ...418L...5P, 1997ApJ...476..232M, Sari_1998}. However,
the high-energy emission is too bright to be explained within the framework of this
basic synchrotron model \citep{1998ApJ...501..772P, 2005ApJ...633.1018P}. For this reason,
it is required to extend the basic afterglow model based
on electron synchrotron to explain the emission at the highest observed energies. 

Various theoretical models were proposed to explain the observed signal at the
high-energy end of the spectrum. A leading suggestion is the synchrotron
self-Compton (hereinafter SSC) mechanism, where the low-energy synchrotron photons
inverse-Compton scatter on the electrons that previously emitted them
\citep{ ghisellini1998quasi, Dermer_1999ApJ, Dermer_2000, 2001ApJ...548..787S, 2009ApJ...703..675N, liu2013interpretation, 10.1093/mnras/stw1175, fraija2019synchrotron, Derishev:2021ivd}. 
Alternatively, it was suggested that the high and very-high energy emission might
be produced by {{accelerated}} baryons. Relativistic hadronic-induced emission
processes such as proton-synchrotron,
photo-pion and photo-pair processes and emission from their secondaries were previously suggested
as viable candidates; see \textit{e.g.} \cite{1998ApJ...499L.131B, 2009ApJ...699..953A, 2010OAJ.....3..150R, 2022arXiv220901940G}.

Direct indication of proton acceleration to high energies is given by the 
explicit detection of high-energy cosmic-rays \citep{asakimori1998cosmic,
sanuki2000precise, lipari2020shape} and of astrophysical PeV
neutrinos by Icecube \citep{2013PhRvL.111b1103, 2015MNRAS.453..113, 2017ARNPS..67...45}.
Their potential association with blazars \citep{PhysRevD.99.103006, 2012MNRAS.426..462P} indicates that relativistic jets might harbor 
high-energy protons, further supporting hybrid and hadronic models for the
high-energy emission; see \textit{e.g.} \cite{2001APh....15..121M, 2002MNRAS.332..215A,
2003APh....18..593M, 2022MNRAS.509.2102G}. Such high energy protons 
could also be present in GRB jets, emitting high energy synchrotron photons \citep{PhysRevLett.78.4328, 1998ApJ...totania, 2001ApJ...559..110Z, 2007MNRAS.380...78G, 2010OAJ.....3..150R}. As we show here, the spectral theoretical expectations from the proton-synchrotron model are consistent with the collective data available for GRB 190114C. We therefore suggest that proton synchrotron may be the leading radiative emission that shape the spectra at the very high ($\sim$ TeV) emission range.  

This paper is organised as follows. In section \ref{sec:Model} we briefly review the hydrodynamical properties of the gas behind the relativistically propagating shock, within the assumptions set by the fireball evolution framework. We then recall the dependence of the Lorentz factor, $\Gamma$, the radius, $r$ and the magnetic field, $B$ in terms of the independent parameters.
Following that, we describe the characteristics of the synchrotron radiation model of the relativistic electrons and protons in section  \ref{sec:sync}. 
Constraints on the model parameters are set from the observations of the afterglow of GRB 190114C in two discrete time intervals in section \ref{sec:Constraints}. 
In section \ref{sec:RESULTS}, we present our results and display the corresponding spectral energy distribution functions under different assumptions. In particular, we explore three special cases in subsections \ref{sec:case1},
\ref{sec:case2} and \ref{best fit}, all linked to uncertainty regarding the characteristics of the particle acceleration mechanism. We discuss contribution of various addition radiative processes in section \ref{sec:5}, and show that they are sub-dominant for the parameters we consider. In section \ref{sec:discussion}, we discuss our results in light of previous works. The conclusions follow in section \ref{sec:Conclusion}.

\section{Afterglow model: dynamics and emission} \label{sec:Model}

\subsection{Dynamics}

We adopt the standard fireball model scenario \citep{1986ApJ...308L..43P,  1986ApJ...308L..47G, 1990ApJ...348..485P,Mesazoras1993ApJ...405..278M, Piran:1993b, 1998ApJ...499..301M}, in which the afterglow emission is explained by the interaction between a highly relativistic jet and its surrounding medium \citep{1993ApJ...418L...5P, Sari_1998}. 
We hereby focus on the self-similar expansion phase, which is expected after a few to a few tens of seconds, during which a relativistically expanding forward shock collects and heats the material from the surrounding interstellar medium (ISM). This heated material, in turn, radiates thereby produces the observed afterglow signal \citep{10.1093/mnras/263.4.861, 1995ApJ...455L.143S, Sari_1998, 2000ApJ...543...66P}. 

When an expanding shell of ultra-relativistic plasma having a Lorentz factor $\Gamma$
interacts with the cold ISM (assumed here to have a  constant density $n$) at rest,
a collisionless shock is formed, and propagates into the ISM.
As the cold ISM material crosses the shock front, its (comoving) density increases to  $4\Gamma n$ and its internal (comoving) energy density becomes $e = 4 \Gamma^2 n m_p c^2$, where $m_p$ is the proton mass and $c$ is the speed of light. 
The time evolution of the Lorentz factor and radius of the shock, $\Gamma$ and $r$, respectively, are given by the self-similar solution of a relativistic blast wave expansion  derived by \cite{1976PhFl...19.1130B},
\begin{align}
   \Gamma (E, n; t) &= \left[ \frac{17 E (1+z)}{1024 \pi n m_p c^5 t^{3}} \right]^{1/8} = 9.4 \hspace{0.3em} E_{53}^{1/8} t_{day}^{-3/8} n_0^{-1/8}, \\
    r(E,n; t) &= \left[\frac{17 E t }{4 \pi n m_p c (1+z)}\right]^{1/4} = 6.4 \times 10 ^{17}\hspace{0.3em} E_{53}^{1/4} t_{day}^{1/4} n_0^{-1/4} \quad\text{cm}, 
\end{align}
where $n = 1 \times n_0 \hspace{0.3em} \text{cm}^{-3}$, $t_{day}$ is the observed
time in days, and the isotropic equivalent kinetic energy is $E = 10^{53} E_{53} \hspace{0.3em} \text{erg}$. Hereinafter any quantity $X_n$ is such that $X = 10^n X_n$ and cgs units aref adopted. In addition, here and in all numerical expressions below, we
consider the redshift to be $z=0.4245$, \textit{i.e.} the observed cosmological redshift of GRB 190114C \citep{web:nasa:stats, web:nasa:23708}. The comoving shell-expansion time (the dynamical timescale) is $t_{dyn} \cong \Gamma t$.

We further use the standard assumption that the magnetic field is generated by the shock, and carries an unknown fraction  $\epsilon_B$ of the internal energy density behind the shock front. As a consequence, in the co-moving frame, the magnetic field strength is 
\begin{align}
   B &= \sqrt{32\pi \epsilon_B \Gamma^2 n m_p c^2} =  0.4 \hspace{0.3em} E_{53}^{1/8} \epsilon_{B,-2}^{1/2} t_{day}^{-3/8} n_0^{3/8}\text{~G}, 
\end{align}

where we took $\epsilon_B = 10^{-2}\epsilon_{B, -2}$.

\subsection{Synchrotron Emission Process} \label{sec:sync}

We assume that both electrons and protons are accelerated by the propagating shock wave, and attain a power-law  distribution $N(\gamma) d\gamma \propto \gamma^{-p} d\gamma$ in the range $\gamma_{\min}\leq \gamma \leq \gamma_{\max}$ as they reach the downstream region. Here, $p$ is the post-shock spectral index which we consider is similar for both the proton and electron populations. Typically, $p$ is assumed to be $p>2$, which is in agreement with both theoretical \citep{1996ApJ...473..204S, PhysRevLett.80.3911, 2000ApJ...542..235K, 2011ApJ...726...75S} and observational analysis of GRB afterglows \citep{2006MNRAS.371.1441S}.   
Let $\gamma_p$ ($\gamma_e$) be the Lorentz factor of a single proton (electron). As the particles radiate, they cool. The radiative cooling time of a particle $x$ (standing for protons or electrons) having Lorentz factor $\gamma_x$ is 
\begin{align} 
   {t_{x, \rm cool}= \frac{\gamma_x m_x c^2}{\frac{4}{3}\sigma_{T,x} c \gamma_x^2 \frac{B^2}{8\pi}}  =\frac{ 6\pi m_x c}{\sigma_{T,x} \gamma_x B^2},} \label{eq:synch_cooling}
\end{align}
   {where $\sigma_{T,p} = (m_e^2/ m_p^2)\sigma_{T,e}$, $\sigma_{T,e}$ is the Thomson cross-section \citep{george1979radiative}. } 
The energy dependence of the cooling time implies the existence of a characteristic Lorentz factor, denoted by $\gamma_c$, for which the cooling time is equal to the dynamical time. 
There are two categories of solution defined by the values of $\gamma_{\min}$ and $\gamma_{c}$: the fast cooling regime, where $\gamma_c < \gamma_{\min}$, and the slow cooling regime, where $\gamma_{\min} < \gamma_c$.

To find the value of $\gamma_{\min}$, we assume that a fraction $\xi_p$ ($\xi_e$) of the proton (electron) population is accelerated and injected to a power law distribution.  
We further assume that the accelerated protons (electrons) carry a fraction $\epsilon_p$ ($\epsilon_e$) of the available energy released as internal energy by the shock.  
Since the average energy per particle in the downstream region is $\Gamma m_p c^2$, and as the average Lorentz factor of an energetic (power-law distributed)
particle is $\bar{\gamma} =  [({p-1})/({p-2})] \gamma_{\min}$ (assuming $p>2$)\footnote{   {Since for $p=2$, the 
coefficient $f(p)$ becomes independent of $p$ and so $f = \ln \left({\gamma_{\min}}/{\gamma_{\max}}\right)$.} }, one finds the minimum injection Lorentz factor of the accelerated
protons and electrons to be 
\begin{align}
  {\gamma_{x, \min} = f(p) \left[\frac{m_p}{m_x}\right]\left[\frac{\epsilon_x}{\xi_x}\right] \Gamma =
  \sim 2 \times 10^3 ~ (\sim 1) ~ 
   f(p) \xi_x^{-1}E_{53}^{1/8} n_0^{-1/8} t_{day}^{-3/8} \epsilon_{x,-1}, } 
\end{align}
   { where the first number refers to electrons and the number in parenthesis refers to protons, $f(p) \equiv (p-2)/(p-1)$  \citep{1996ApJ...473..204S, Sari_1998, Granot_2002}}. Here and below, we consider the number density of the accelerated electrons, $n_e = \xi_e n$ and of the accelerated protons,  $n_p = \xi_p n$.

By balancing the co-moving dynamical time
$t_{dyn}$ and the cooling time by synchrotron emission $t_{cool}$, the cooling
Lorentz factors of the particle species are $ \gamma_{p,c} = ({6\pi m_p^3 c})/({\sigma_Tm_e^2B^2\Gamma t})$ 
and $\gamma_{e,c} = ({m_e}/{m_p})^3\gamma_{p,c}$.
A third characteristic Lorentz factor is the maximum Lorentz factor achieved by the particles \citep{1996ApJ...457..253D}. It can be estimated as follows. As the particles cool by the synchrotron process, their maximum achievable Lorentz factors are
calculated by equating the energy loss time and the particle
acceleration time, $t_{\rm acc} = 2 \pi \alpha \epsilon/qcB $
\footnote{Note that $\epsilon/ (cqB) = r_L$ is the
Larmor radius. Here, the acceleration time $t_{\rm acc}$ is for Fermi-type accelerations.} 
where $\epsilon = \gamma m c^2$ is the particle's energy, and $\alpha \geq 1$ is a numerical coefficient {normalising the acceleration time to the Larmor time}. The
maximum Lorentz factor is thus
\begin{align}
  {\gamma_{x,\max} = \left[\frac{6\pi q}{\alpha \sigma_{T,x} B} \right]^{1/2} = \sim 2 \times 10^8~ (3.6 \times 10^{11})~
   \alpha^{-1/2} E_{53}^{-1/16} n_0^{-3/16} t_{day}^{3/16} \epsilon_{B,-2}^{-1/4}.}
   \label{eq:proton_gamma_max}
\end{align}
In principle the coefficient $\alpha$ for electrons can differ from that of the protons. In our analysis, the only constraining one is that of the protons.
For the parameters of the model presented here, a large magnetic field is required and therefore the electrons are in the fast cooling regime with $\gamma_{\min} > \gamma_c$. The resulting electron distribution is a broken power-law with index 2 between $\gamma_c$ and $\gamma_{\min}$ and index $p$ between $\gamma_{\min}$ and $\gamma_{\max}$. On the other hand, the protons are in the slow cooling regime. Therefore their distribution function is a single power-law with index $(p-1)$ between $\gamma_{p,  \min}$ and $\gamma_{p, \max}$ (for the parameters we find, $\gamma_{p, \max} < \gamma_{p, c}$, see below).

The observed synchrotron spectra from these particle distributions have the broken-power law shape with three characteristics frequencies \citep{1970RvMP...42..237B, 1999ApJ...511..852G,Granot:2000tta, Granot_2002}. The observed characteristics spectral peak frequency of photons emitted by the particles at $\gamma_{\min}$ is 
\begin{equation} \label{equ:vmin}
    {\nu_{\min} = \frac{3}{4\pi}\frac{q B}{m_x c}\gamma_{\min}^2 \left[\frac{\Gamma}{1+z}\right] =\sim 
    0.124  ~(2.07 \times  10^{-11})~f(p)^2 E_{53}^{1/2} t_{day}^{-3/2} \epsilon_{B,-2}^{1/2}\epsilon_{x,-1}^2 \xi_x^{-2} \quad \text{eV}.}
\end{equation}
The cooling frequencies of the electrons and of the protons are
\begin{equation} \label{equ:vc}
    {\nu_{c} = \frac{3}{4\pi}\frac{q B}{m_x c}\gamma_{c}^2
        \left[\frac{\Gamma}{1+z}\right] =\sim 2.19 \times 10^{-3} (4.55 \times 10^{13})~
    E_{53}^{-1/2} t_{day}^{-1/2} \epsilon_{B,-2}^{-3/2} n_0^{-1} \quad \text{keV}.}
\end{equation} 
Note that the proton cooling frequency $\nu_{p,c}$ is $\sim 10^{16}$ times larger
than the electron cooling frequency $\nu_{e,c}$. {As a result of their higher mass, protons cool slower than electrons.}

The maximum synchrotron frequencies for electrons and protons are
\begin{equation} \label{equ:vmax}
       {\nu_{\max} =\frac{3}{4\pi}\frac{q B}{m_x c}\gamma_{\max}^2
     \left[\frac{\Gamma}{1+z}\right] =\sim
     1.65 \times 10^3 (2.89\times10^{6})~ \alpha^{-1} E_{53}^{1/8}n_{0}^{-1/8}t_{day}^{-3/8} \quad \text{MeV}.}
\end{equation}
Consequently, for fiducial values of the parameters, the highest energy of photons produced by the electron-synchrotron process is limited to $\leq$ few GeV. Therefore, photons having energies much higher than that must have a different origin. 
From the previous equation, the cooling and injection frequencies are similar. However, a model in which proton synchrotron explains the TeV  observations requires a large magnetic field, which translates into a fast cooling regime for the electron $\nu_{e,c} \ll \nu_{e,\min}$. However, for protons, $\nu_{p,c} \gg \nu_{p,\min}$ implying that the protons are in the slow cooling regime, and lose their energy inefficiently. {In fact, protons satisfy $\nu_{\rm p,c}>\nu_{\rm p,max}$. Therefore, the proton synchrotron spectrum is a power-law with a high energy cutoff at frequency $\nu_{\rm p,max}$.}

The maximum observed radiative power from a single particle (proton or electron) at observed frequency $\nu_{obs} = \nu_{peak}\Gamma/(1+z)$
is given by
\begin{equation}
P_{\nu_{p,\max}} = (1+z)\frac{2}{9} \frac{m_e^2 c^2 \sigma_T}{q m_p} B\Gamma
\end{equation}
and $P_{\nu_{e,\max}} = ({m_p}/{m_e})P_{\nu_{p,\max}}$ \citep{george1979radiative, 1996ApJ...473..204S, Sari_1998}.
\begin{equation}\label{equ:Fv}
     {F_{\nu_{x,\rm peak}} =\frac{N_x P_{\nu_{x,\max}}}{4\pi d_L^2} =\sim
     {100 (0.06)~E_{53} \xi_x \epsilon_{B,-2}^{1/2}n_0^{1/2} d_{L, 28}^{-2} \quad \text{mJy}.}}
\end{equation} 
where $N_x = (4\pi/3) n_x r^3 $ is the number of particles swept by the blast-wave which are actively radiating, where $n_x = \xi_x n$. Here, $d_{L}$ is the luminosity distance. 

For particles accelerated to a power law, $N(\gamma) d\gamma \propto \gamma^{-p}$ above $\gamma_{\min}$ and below $\gamma_{\max}$, the expected photon spectrum thus is a broken power-law shape, with $F_\nu \propto \left\{ \nu^{1/3}, \nu^{-(p-1)/2}, \nu^{-p/2} \right\}$ for $\left\{ \nu < \nu_{\min}, \nu_{\min} < \nu < \nu_c, {\nu_c < \nu < \nu_{\max}} \right\}$ in the slow cooling regime and $F_\nu \propto \left\{ \nu^{1/3}, \nu^{-1/2}, \nu^{-p/2} \right\}$ for $\left\{ \nu < \nu_{c}, \nu_{c} < \nu < \nu_{\min}, {\nu_{\min} < \nu < \nu_{\max}} \right\}$ in the fast cooling regime \citep{george1979radiative, Sari_1998}.

\section{Model constraints derived from the available data of  GRB 190114C} \label{sec:Constraints}

We proceed to interpret the available data of GRB 190114C
within the framework of a hybrid model,
for which the low energy component is explained by synchrotron
radiation from electrons while the high energy TeV component
is required to be proton synchrotron. GRB 190114C was a long-GRB with a prompt energy released $E_{\rm iso} \simeq 2.5 \times 10^{53}$~erg
\citep{2020ApJ...890....9A}. It is seen at the
redshift $z=0.4245$; identified by the
Nordic Optical Telescope \citep{web:nasa:stats} and further established by Gran
Telescopio Canarias \citep{web:nasa:23708}. 
Its prompt phase was recorded over an energy band of $8$~keV - $100$~GeV
by the \textit{Swift}-Burst Alert Telescope (BAT) \citep{web:nasa:swift}, the Gamma-ray Burst Monitor
(GBM) \citep{web:nasa:Fermi}, and the Large Area Telescope (LAT) \citep{web:nasa:FermiLAT}. The reported duration is $T_{90} = 25$~s \citep{web:nasa:swift, MAGIC:2019irs}, although the GBM collaboration reported a duration of $T_{90} = 116$~s \citep{web:nasa:Fermi}. This longer duration is explained by the observation of a weak second emission episode after the initial signal. This episode was interpreted as emission from the afterglow \citep{2020ApJ...890....9A}, and in this paper we make the same assumption. 
 
The afterglow  follow-up observations were carried out by many instruments. As a
result a good temporal and multi-wavelength data set 
exists during the early afterglow phase when the burst was bright enough in the TeV band to allow the MAGIC instrument to
measure the spectrum in five time bins; 68 - 110 s, 110 - 180 s, 180 - 360
s, 360 - 625 s and 625 - 2400 s \citep{MAGIC:2019irs}. The first two time intervals
have a spectral data in the keV to  GeV band (XRT, GBM, LAT) along with
data in the TeV band from MAGIC.
Inspection of the available data reveals that the afterglow spectra during the first
two time bins centered at $90$~s 
and $120$~s are characterized by two peaks, with the
lowest peak at
energy around 10 keV which we attribute to synchrotron emission
from electrons, and a second
peak 
between the GeV and the TeV bands which we interpret as synchrotron emission from
protons. In this section, we derive the relations between the model parameters to satisfy these two assumptions. 

According to the XRT online repository \citep{2009MNRAS.397.1177E}, the 
X-ray spectrum of GRB 190114C is fitted by an absorbed power-law, characterised by a photon index of $1.7^{\pm
0.04}$. With the XRT online tool \citep{2009MNRAS.397.1177E}, we checked that this holds in the first and second time bins independently. We find the photon index to be $1.68^{+0.12}_{-0.11}$ between 68 and 110s, and $1.58^{\pm 0.1}$ between 110 and 180s. This power-law can naturally be produced in the electron-synchrotron model. Since for
the fiducial values of the model parameters discussed herein (see discussion above) electrons are expected to be in the fast cooling regime,
the spectral slope can be obtained in two cases. The first one
corresponds to (i) $\nu_c < \nu_{\rm XRT} < \nu_{\min}$, producing a
spectral index of $-1.5$ comparable to the spectral index found from
the XRT repository. In this case the electron index cannot be
determined and we can take $p_e \sim 2.2$. We discuss the impact of
this assumption below. In the other case, $\nu_c < \nu_{\min} < \nu_{\rm XRT} < \nu_{\rm max}$, the electron index is deduced to be $p_e = 1.2 $ as the observed photon spectral index is 1.6. It also
requires  $\nu_{\rm max}$ to be close to 10~keV to produce the first
hump. However, inspection of Equation \eqref{equ:vmax} reveals that
this case is difficult to achieve as it requires the observed time, $t_{day}$ to be long
(which is in contrast with the observation at hundreds seconds) and a large $\alpha$, { meaning that the constraints on the particle acceleration mechanism are loose}. 

The available multi-wavelength data are now used to constrain the free model
parameters. For this purpose, we use the available data  at the two time bins
centered at $90$~s and $120$~s.
\citet{2020ApJ...890....9A}  reported a break at 4.72 keV (at 68 - 110s) and  5.6 keV (at 110 - 180s), and therefore
we assume that the injection frequency
$\nu_{e,\min}$ is equal to 5.5 keV consistent with those findings.  The large magnetic field requirement in our model implies that the cooling frequency $\nu_{e,c}$ 
to be below the XRT band, \textit{i.e.} smaller than 0.3 keV. Using Equation \eqref{equ:vmin} for the value of the injection frequency yields 

\begin{equation}
E_{53}^{1/2}\epsilon_{B,-2}^{1/2} \epsilon_{e,-1}^2 \xi_e^{-2} = \left \{
\begin{aligned} \label{equ: v_min limit}
 &  53.65   & ~~~~~~ &{\rm at~}  t=90\hspace{0.3em}{\rm s},\\
 &  82.60   & & {\rm at~} t=120\hspace{0.3em}{\rm s},
\end{aligned}  \right.
\end{equation}
while using Equation \eqref{equ:vc} for the cooling frequency gives
\begin{equation} \label{equ: v_c limit}
E_{53}^{-1/2}\epsilon_{B,-2}^{-3/2} n_0^{-1} \leq \left \{
\begin{aligned}
& 4.42 & ~~~~~~ &{\rm at~} t=90\hspace{0.3em}{\rm s},\\
& 5.10  & ~~~~~~ &{\rm at~} t=120\hspace{0.3em}{\rm s}.
\end{aligned}  \right.
\end{equation}
   {The maximum observed photon energies, $\sim 1$~TeV (90 s and 120 s) \citep{MAGIC:2019irs}}, can be used to set an estimation of the maximum frequencies of photons radiated from the protons, see Equation \eqref{equ:vmax}. {Using these observed maximum energies at 90 and 120s, we constrain the parameters.} We get : 
\begin{equation} \label{equ: v_maxp limit}
   {\frac{1}{\alpha} \left ( \frac{E_{53}} {n_0} \right ) ^{1/8} = \left \{
\begin{aligned}
& 2.63 \times 10^{-2} & ~~~~~~ &{\rm at~}  t=90\hspace{0.3em}{\rm s},\\
& 2.93  \times 10^{-2} & ~~~~~~ &{\rm at~}  t=120\hspace{0.3em}{\rm s}. 
\end{aligned}    \right.}
\end{equation}

We now use the specific fluxes at $\nu_{\min}$ and $\nu_{p, \max}$ to constrain the parameters. We first assume a power-law index $p=2.2$ for both electrons and protons, 
the fluxes corresponding to the observed energies at the time period of 90 s ($F_{\nu_{obs} =5.5
~\rm keV} \approx 2.56 \times
10^{-26}$ ergs cm$^{-2}$ s$^{-1}$ Hz$^{-1}$ and $F_{\nu_{obs}=0.23~\rm TeV} \approx 9.5 \times
10^{-34}$ ergs cm$^{-2}$ s$^{-1}$ Hz$^{-1}$) and 120 s ($F_{\nu_{obs} = 5.5~\rm keV} \approx 1.41 \times 10^{-26}$ ergs cm$^{-2}$ s$^{-1}$ Hz$^{-1}$ and $F_{\nu_{obs} = 0.23~\rm TeV} \approx4.15 \times 10^{-34}$ ergs cm$^{-2}$ s$^{-1}$ Hz$^{-1}$), 
can be explained using the synchrotron
fluxes ($F_{\nu_e}$ and $F_{\nu_p}$) calculated from Equations
\eqref{equ:vmin}, \eqref{equ:vc} and \eqref{equ:Fv}. Further using Equation \eqref{equ: v_min limit} provides
restrictions on the parameter space as follows,
\begin{equation} \label{equ: Fve}
E_{53}^{3/4} \epsilon_{B,-2}^{-1/4} \xi_e \approx \left \{
\begin{aligned}
&0.21 & ~~~~~~ &{\rm for~}  t=90\hspace{0.3em}{\rm s},\\
&0.12 & ~~~~~~ &{\rm for~}   t=120\hspace{0.3em}{\rm s}. 
\end{aligned} \right.
\end{equation}
and
\begin{equation} \label{equ: Fvp}
 E_{53}^{13/10} \epsilon_{B,-2}^{4/5} n_0^{1/2} \epsilon_{p,-1}^{6/5}\xi_p^{-1/5} \approx \left \{
\begin{aligned}
& 2.4 \times 10^{5} & ~~~~~~ &{\rm for~}  t=90\hspace{0.3em}{\rm s},\\
& 1.3 \times 10^{5} & ~~~~~~ &{\rm for~}   t=120\hspace{0.3em}{\rm s}. 
\end{aligned} \right.
\end{equation}
Equation \eqref{equ: Fve} is derived from the relation $F_{\nu_e} = F_{\nu_e,peak} (\nu_{e,\min}/\nu_{e,c})^{-1/2} (\nu/\nu_{e,\min})^{-p/2}$ for the synchrotron emission of the fast-cooling electrons and Equation \eqref{equ: Fvp} is obtained from $F_{\nu_p} = F_{\nu_p,peak} (\nu/\nu_{p,\min})^{-(p-1)/2}$ relevant for the synchrotron emission from the protons. 

Using those two last equations and assuming the value $p = 2.2,$ one can express $\epsilon_B$ and $\xi_e$ as,

\begin{equation} \label{eq:epsilonB}
\epsilon_{B,-2} 
  = \Biggl \{ \begin{array}{ll}
5.3 \times 10^6  \\ 2.5 \times 10^6 
\end{array}
\xi_p^{1/4}E_{53}^{-13/8} \epsilon_{p, -1}^{-3/2} n_0^{-5/8} ~~~~~~
 \begin{array}{ll}
        {\rm for~}  t=90\hspace{0.3em}{\rm s} &  \\
        {\rm for~}  t=120\hspace{0.3em}{\rm s} & 
    \end{array}
\end{equation}
and for $\xi_e$:
\begin{equation} \label{eq:xi_e}
 \xi_e =
  \Biggl \{
\begin{array}{ll}
  &10.08 \\ &4.76
  \end{array}  \xi_p^{ 1/16} E_{53}^{-37/32} \epsilon_{p, -1}^{-3/8} n_0^{-5/32} ~~~~~~ 
 \begin{array}{ll}
        {\rm for~}  t=90\hspace{0.3em}{\rm s} &  \\
        {\rm for~}  t=120\hspace{0.3em}{\rm s} & 
 \end{array}
\end{equation}
We further use Equations \eqref{eq:epsilonB} and \eqref{eq:xi_e} with Equation \eqref{equ: v_min limit} to obtain the value of $\epsilon_e$ :
\begin{equation}\label{eq:epsilone}
   \epsilon_{e, -1} =
    \Biggl \{
    \begin{array}{ll}
     1.54 \\  1.09 
     \end{array}  E_{53}^{-1} ~~~~~~ 
    \begin{array}{ll}
        {\rm for~}  t=90\hspace{0.3em}{\rm s} &  \\
        {\rm for~}  t=120\hspace{0.3em}{\rm s} & 
     \end{array}
\end{equation}

In our analysis below, we also use the physical condition that the electrons, protons and magnetic field energy are all obtained from the post-shock thermal energy, namely $\epsilon_B+\epsilon_e+\epsilon_p\leq 1$. Inspections of those equations reveals that the time bin centered at 90 s gives the adequate parameter magnitudes. Therefore, hereinafter we will use 
the constraints obtain from this time bin only and we will show that the time evolution can be well reproduced by the afterglow dynamics alone.

It is clear that satisfying the condition  $\epsilon_B < 1$ requires a combination of (i) a small $\xi_p$  \textit{i.e.} a small fraction of protons being accelerated into a power-law, (ii) a large total kinetic energy $E_{53}$, (iii) a large circumburst medium density $n_0$ and (iv) a large (but smaller than the unity) $\epsilon_p$, \textit{i.e.} a large fraction of energy given to accelerated protons. We note that this combination of parameters also leads to a small value of $\xi_e$, namely, only a small fraction of the electron population is accelerated to a power law. 

\begin{figure}[t!]
     \centering
     \begin{tabular}{ccc}
        \includegraphics[width=0.31\textwidth]{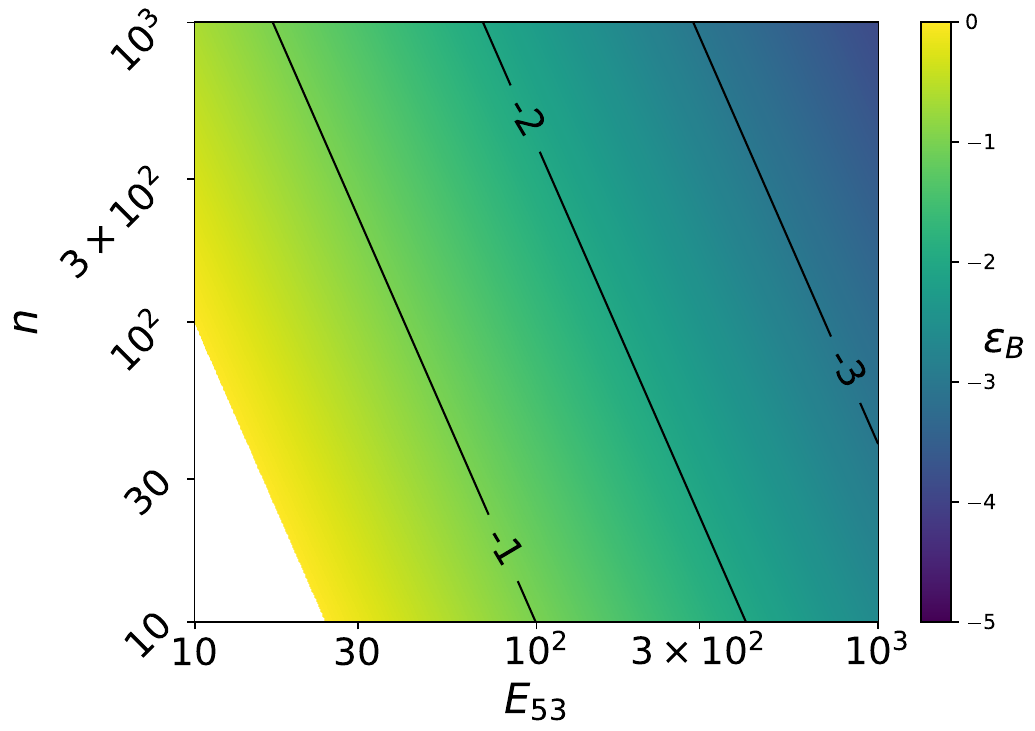} &
        \includegraphics[width=0.31\textwidth]{Figures/epsilon_B_for_xi_p_=_0.01.pdf} &
        \includegraphics[width=0.31\textwidth]{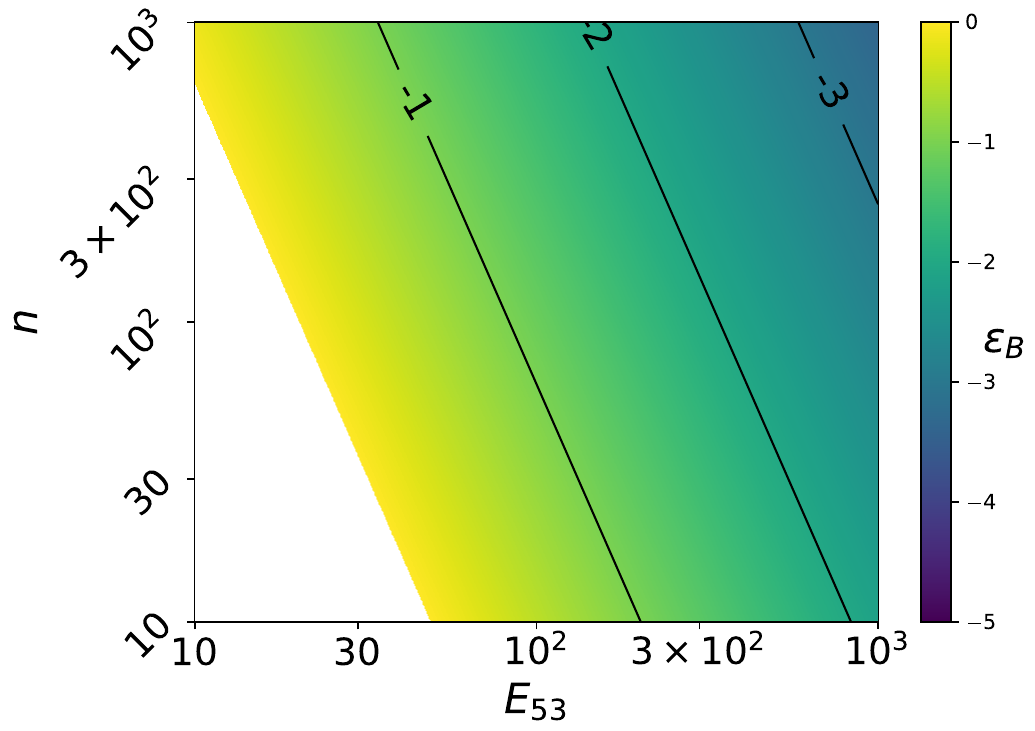} \\
        $\xi_p = 0.01$ & $\xi_p = 0.1$ & $\xi_p = 1$ 
     \end{tabular}
        \caption{ Colors give the value of the magnetization parameter $\epsilon_B$ (in logarithmic scale) as a function of the ambient density, $n_0$ and total energy, $E_{53}$ from Equation \eqref{eq:epsilonB} for the time bin centered at 90~s. In making the plot, we assumed $\epsilon_p = 0.8$ and display the results for three different values of $\xi_p$, namely, $\xi_p = 10^{-2},10^{-1},1$, respectively presented from left to right. The thin black lines represent combination of parameters that result in $\epsilon_B = 10^{-1}, 10^{-2}, 10^{-3}$.}
        \label{fig:epsilon_B plot}
\end{figure}

We show the constraints on the parameters in Figure \ref{fig:epsilon_B plot}, which displays the value of $\epsilon_B$ as a function of $n$, $E_{53}$ and $\alpha$. Inspection of Equation \eqref{equ: v_min limit} reveals that the dependence of $\epsilon_e$ on $E_{53}, n_0$~and $\epsilon_p$, implies that the condition $\epsilon_e < 1$ is automatically satisfied, when satisfying the limitation given by Equation \eqref{eq:epsilonB}. Finally, inspecting Equation \eqref{equ: v_maxp limit}, it comes that the dependence on $E_{53}$ and $n_0$ requires $\alpha$ to not be too large, with $\alpha \sim 100$ owning for the weak dependence on $E_{53}$ and $n_0$.

\section{Results : Explaining the TeV observation of GRB 190114C with proton synchrotron}\label{sec:RESULTS}

In this section, we present several solutions for a proton synchrotron model
to explain the TeV emission of GRB 190114C. In particular we consider three assumptions
based on the uncertainty of the particle acceleration process: (i) similar particle
spectral indices, $p_e = p_p$ and similar fraction of accelerated particles, $\xi_e
= \xi_p$ \textit{i.e} the acceleration process is similar for protons and electrons
in terms of number of accelerated particles and the obtained spectral shape. (ii)
Similar particle spectral indices, $p_e = p_p$ but $\xi_e \neq \xi_p$, namely the
acceleration process accelerates different proportions of electrons and protons, but
produces a similar spectral shape. And (iii) different spectral indices, $p_e \neq
p_p$ with $\xi_e = \xi_p$, meaning that the acceleration process produces a different spectral shape. We find that the effect of $\xi_p$ on the resulting
spectra is not significant, therefore we did not have to assume an extra degree of freedom (see below)\footnote{Since we normalize the flux to the observed flux,
reducing $\xi_p$ enforces an increase in the values of other free model parameters,
such as the magnetization.}.

The results are presented in Figures \ref{fig:xie = xip graphs}, \ref{fig:xie neq xip graphs} and \ref{fig:two graphs} respectively. The data are extracted from \cite{MAGIC:2019irs}
and are presented here for convenience. In particular, we did not aim at producing
the statistical best fit to the data, but only to demonstrate the ability of our
model to reproduce the fluxes and characteristic breaks in the XRT and MAGIC bands.
This allows us to constrain the values of the free parameters of our model.   

\subsection{Case (i): \texorpdfstring{$\xi_e = \xi_p$}{TEXT}} \label{sec:case1}
\begin{figure}[t!]
     \centering
     \begin{tabular}{cc}
        \includegraphics[width=0.45\textwidth]{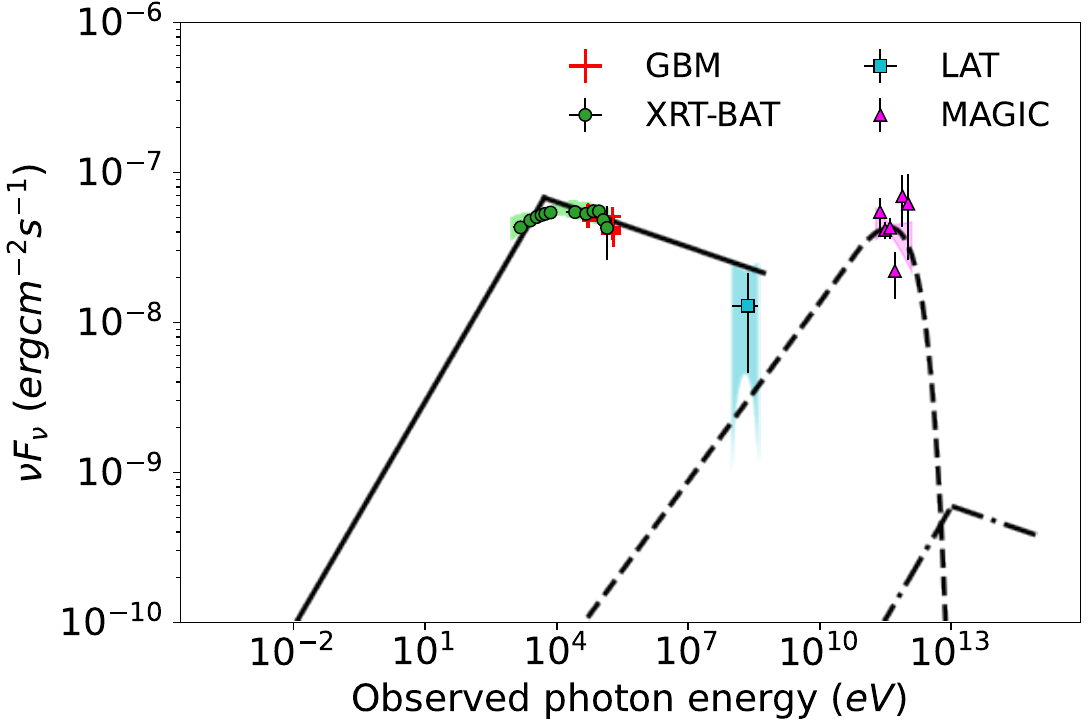} &
        \includegraphics[width=0.45\textwidth]{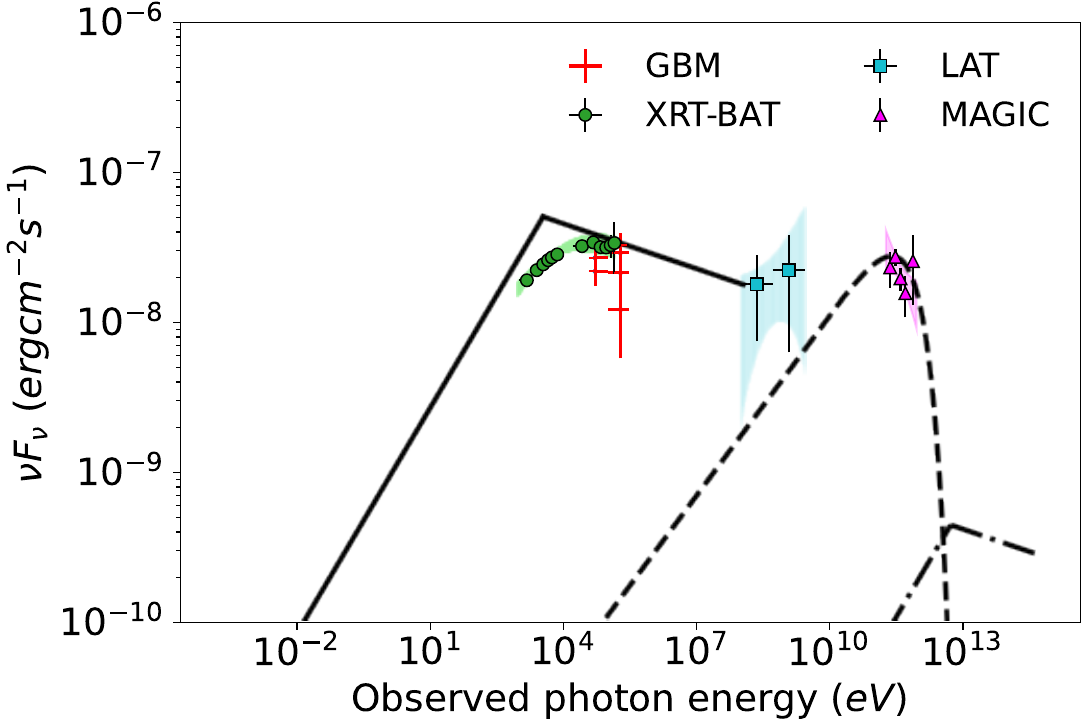} \\
        a) $t=90$~s &  b) $t=120$~s
     \end{tabular}
     \caption{Resulting spectra for parameters in our case (i) : $\xi_e = \xi_p $ plotted on top of the data of GRB 190114C at 90~s and 120~s. The SED obtained from
     our model for the time bins (a) [$\rm T_0 + 68- T_0 + 110$] and (b) [$\rm T_0 +
     110- T_0 + 180$]. Continuous line: electron synchrotron, dashed line: proton
     synchrotron and dotted-dashed line: inverse Compton. The data are from \cite{MAGIC:2019irs}
     and are displayed to guide the eye. The parameters are $E_{53} = 40$, $n_0 = 80$, $\epsilon_{p, -1} = 8$, $p = 2.2$ and $\alpha = 35$, resulting in $\epsilon_{B,-2} = 15.2$, $\xi_e = \xi_p = 0.026$ and $\epsilon_{e,-1} = 0.0385$. The cut-off at $\gamma_{p,\max}$ is attained by $\exp (-\nu/\nu_{\max})$.}
     \label{fig:xie = xip graphs}
\end{figure}

We first examine the  assumption of similarity between the injection fraction of
electrons and protons in the acceleration process. 
We impose $\xi_e =
\xi_p$ in Equation \eqref{eq:xi_e} for $t =90$~s which results in
\begin{align}  \label{eq:xie=xip} 
    {\xi_e \equiv 11.76 \quad E_{53}^{-37/30}\epsilon_{p,-1}^{-6/15} n_0^{-1/6}.}
\end{align}
Also in order for $\epsilon_B$ to be smaller than unity, the kinetic energy is
required to be greater than the observed released prompt energy
$E_{\rm iso}$ as seen explicitly from Equation \eqref{eq:epsilonB}.
Here we choose $E = 4 \times 10^{54}$~ergs, $\epsilon_p = 0.8$, and ambient density $n = 80$~cm$^{-3}$, then Equation \eqref{eq:xie=xip} yields $\xi_e =\xi_p = 0.026$, for which $\epsilon_B = 0.152$ and $\epsilon_e = 3.85 \times 10^{-3}$ and so the relation $\epsilon_B+\epsilon_e+\epsilon_p\leq 1$
is satisfied. For this set of parameters, only $2.6\%$ of the electrons and
protons injected attain the power-law distributions behind the shock front and the rest assume a thermal distribution with temperature lower than $\gamma_{\min}$. We do
not attempt to model the radiation from these thermal particles, as their contribution is below the observed band. 
Equation \eqref{equ: v_maxp limit} constrain the numerical coefficient that determines the acceleration efficiency to be {$\alpha = 35$}.
Above $\gamma_{p, \max}$ we assume an exponential cutoff representing the inability of the acceleration process to accelerate protons to energies above $\gamma_{p, \max}$ (see dashed line in Figure \ref{fig:xie = xip graphs}). 
For this set of parameters, the spectra obtained in the time bins
centered at 90 s and 120 s are displayed in Figure \ref{fig:xie = xip graphs}.
This set of parameters results in a spectra and flux consistent with the observed data in both time intervals. The large value of $E$ implies that the
efficiency $\eta=E_{\rm iso}/(E+E_{\rm iso})$ of the prompt emission $\eta \simeq 5.9\%$ is low, but not extremely low. Similarly, the required energy is large, but acceptable. We comment on the low efficiency in the discussion below in section \ref{sec:discussion}.

\subsection{\texorpdfstring{Case (ii): $\xi_e \neq \xi_p$}{TEXT}} \label{sec:case2}
We next examine a model in which the electrons and protons are
accelerated in different numbers, $\xi_e \neq \xi_p$. The parameters we consider, $E_{53} = 50$,
$n_0 = 130$, $\epsilon_p = 0.8$, {$\alpha = 34$}, and $\xi_p = 0.2$ provides 
adequate spectra that are consistent with the observed data at both 90~s and 120~s. Using Equation \eqref{eq:epsilonB} we obtain 
$\epsilon_B = 0.13$, and from Equation \eqref{eq:xi_e} one finds $\xi_e = 0.02$ and $\epsilon_e =  3.08 \times 10^{-3}$.

\begin{figure}[t!]
     \centering
     \begin{tabular}{cc}
         \includegraphics[width=0.45\textwidth]{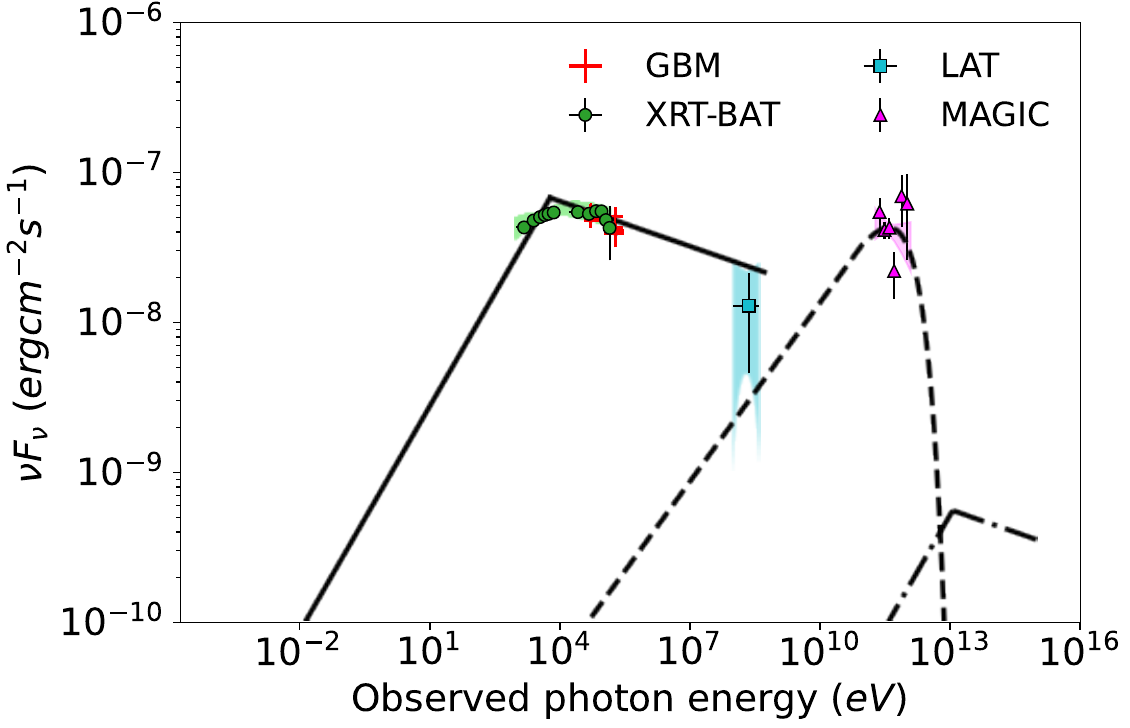} &
         \includegraphics[width=0.45\textwidth]{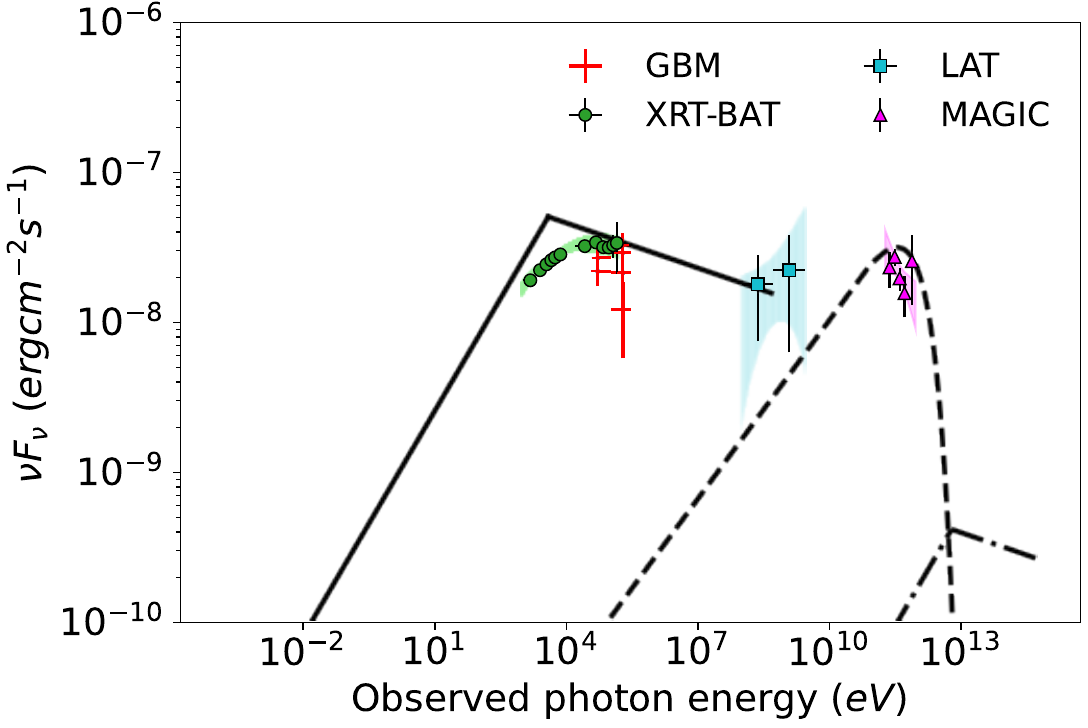} \\
        a) $t=90$~s & b) $t=120$~s \\
     \end{tabular}
     \caption{Case (ii). Same as Figure \ref{fig:xie = xip graphs}. The parameters are $E_{53} = 50$, $n_0 = 130$, $\epsilon_{p, -1} = 8$, $p = 2.2$, $\alpha = 34$ and $\xi_p = 0.2$, $\epsilon_{B,-2} = 13$, $\epsilon_{e,-1} = 0.0308$, $\xi_e = 0.02$ and so $\xi_e/\xi_p \simeq 0.1$. The cut-off at $\gamma_{p,max}$ is attained by $\exp (-\nu/\nu_{\max})$.}
     \label{fig:xie neq xip graphs}
\end{figure}

Our results are presented in Figure
\ref{fig:xie neq xip graphs} for the time intervals centered at 90 s and 120 s.
The parameters of this solution are nearly the same as the ones obtained
for the case $\xi_e = \xi_p$. This is because the protons dominate the
energetic requirements. In particular, it means that the total kinetic energy
of the blast-wave has to be large, and correspondingly, the efficiency is low. In this case it is $\eta \sim 4.8\%$. Owning for the large energy, the interstellar
medium density is also large. Again we point out that all the values we obtain are in range with estimates and uncertainties of GRB energetics and ambient densities.

\subsection{Case (iii) \texorpdfstring{$\xi_e \equiv \xi_p$}{TEXT} and \texorpdfstring{$p_e \neq p_p$}{TEXT}} \label{best fit}

\begin{figure}[t!]
     \centering
     \begin{tabular}{cc}
        \includegraphics[width=0.45\textwidth]{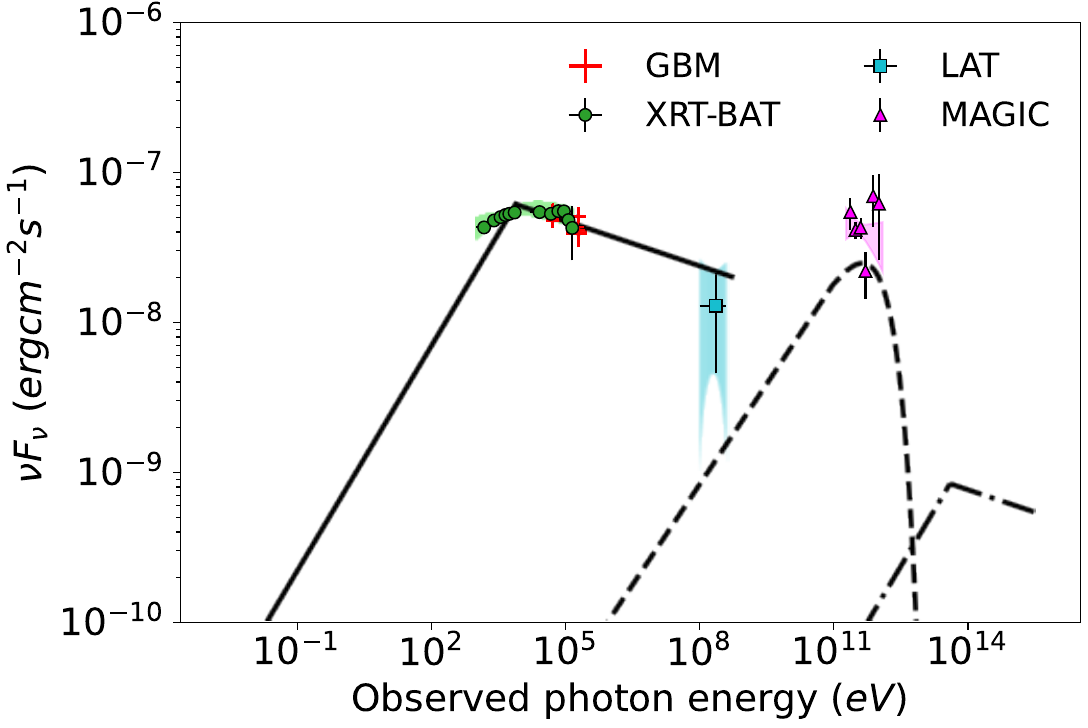} &
        \includegraphics[width=0.45\textwidth]{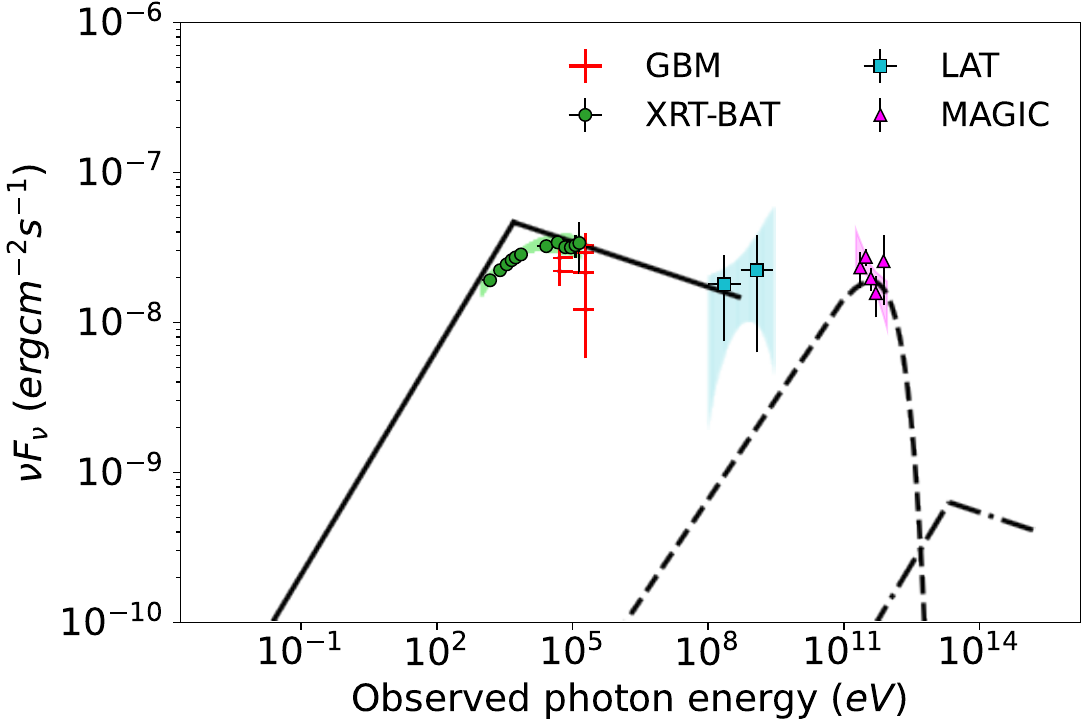} \\
        a) $t=90$~s & b) $t=120$~s
     \end{tabular}
     \caption{Case (iii) : $\xi_e = \xi_p = 0.023$, and $p_e = 2.2$ while $p_p = 2.1$. Same as Figure \ref{fig:xie = xip graphs}. The parameters we used here are $E_{53} = 30$, $n_0 = 15$, $\epsilon_{p,-1} = 8$, and $ \alpha = 41.45$, $\epsilon_{B,-2} = 12$ and $\epsilon_{e,-1} = 0.047$. The cut-off at $\gamma_{p,max}$ is attained by $\exp (-\nu/\nu_{\max})$.}
     \label{fig:two graphs}
\end{figure}

The critical parameter which determines the  energetic budget is
the proton injection index $p_p$. We have assumed the electron index to be 2.2 in agreement with theory \citep{Sari_1998} and observations \citep{2009MNRAS.397.1177E}. The proton index was then assumed to be equal to the electron index. Here we relax 
this assumption. Since the TeV flux is modeled by emission from protons at $\gamma_{p, \max}$, a high value of $p_p$ would require higher energy budget, and lower efficiency. We therefore consider $p_p = 2.1 < p_e$ in order to decrease the energy
budget and increase the efficiency. The solution we present here is given
by the parameters $n = 15$~cm$^{-3}$, $E = 3 \times 10^{54}$~ergs, $\epsilon_p = 0.8$. This gives {$\alpha = 41.45$},
$\epsilon_e = 4.7 \times 10^{-3}$, $\epsilon_B = 0.12$, and $\xi_e = \xi_p = 0.023$. The resulting
spectrum in the two time bins centered at 90 s and 120 s is shown in
Figure \ref{fig:two graphs}.

A consequence of the assumptions used here is that the prompt phase efficiency is increased to $7.7\%$, which is higher than in the two previous cases. Furthermore, the density of the interstellar medium is substantially lower than the value obtained in our previous scenarios. 

\begin{figure}[t!]
     \centering
     \begin{tabular}{ccc}
        \includegraphics[width=0.31\textwidth]{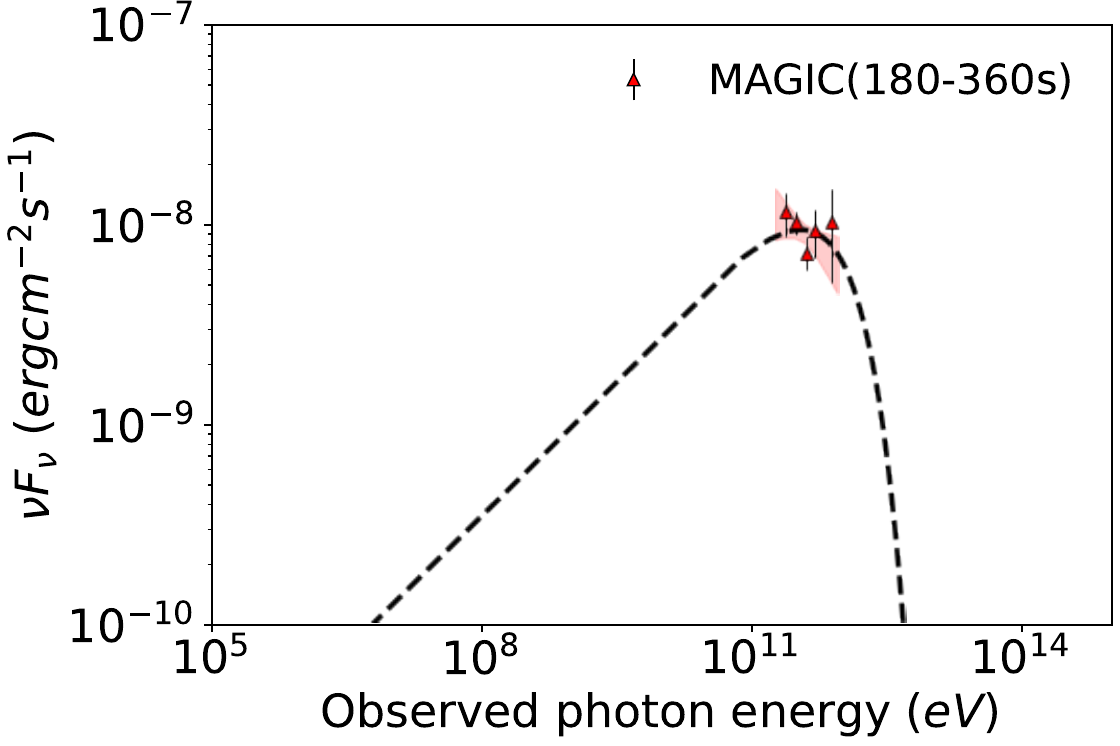} &
        \includegraphics[width=0.31\textwidth]{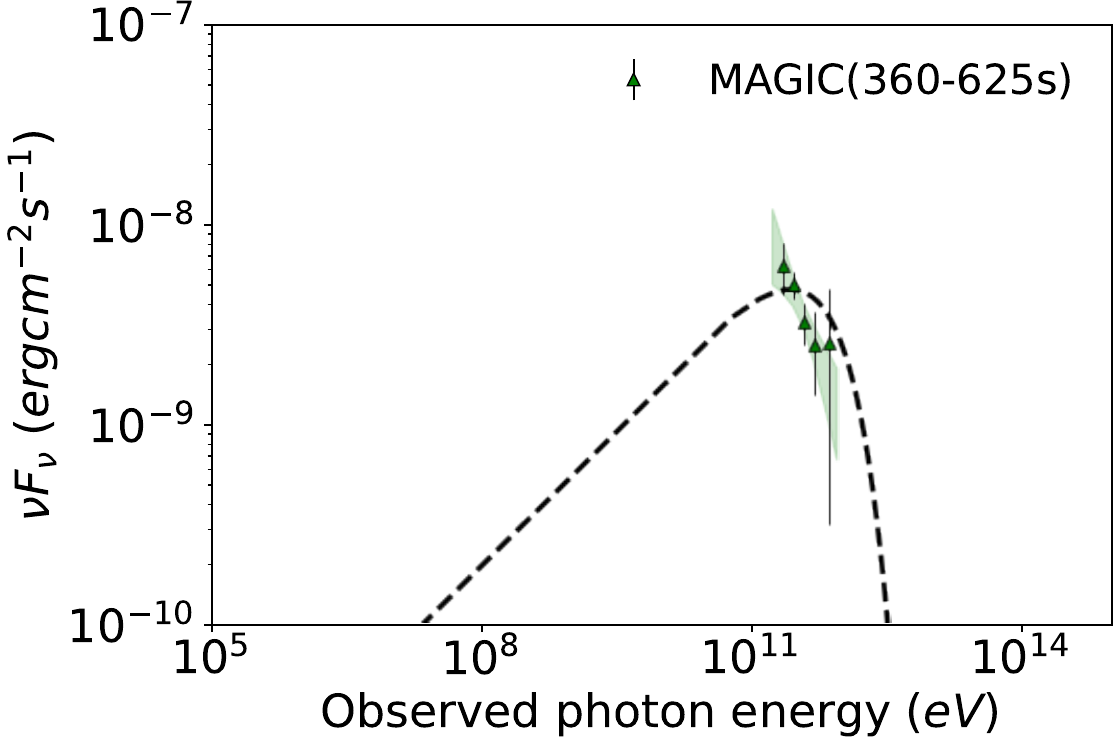} &
        \includegraphics[width=0.31\textwidth]{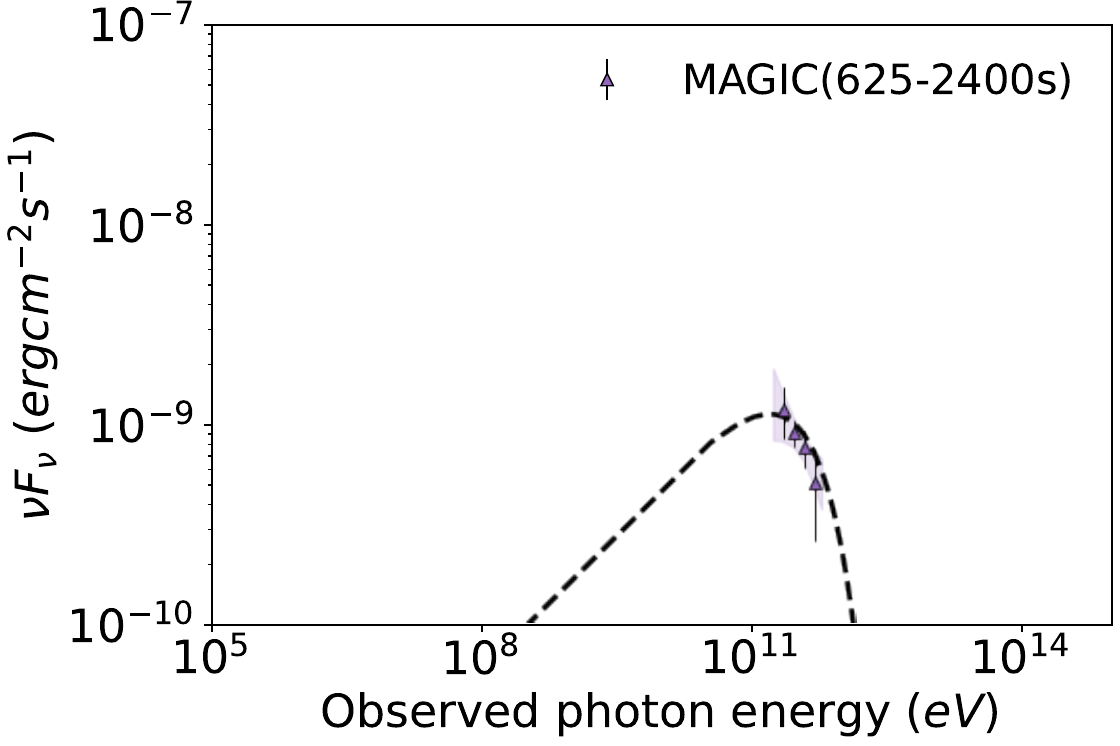} \\
        $t=260$~s &    $t=520$~s & $t=2200$~s
     \end{tabular}
        \caption{The proton synchrotron component for our model in case (iii), together with the data from the three additional MAGIC observations \citep{MAGIC:2019irs}. The depletion of the observed flux is in agreement with our model.}
        \label{fig:three graphs}
\end{figure}

\begin{figure}[ht!]
    \centering
    \includegraphics[width = 0.7\textwidth]{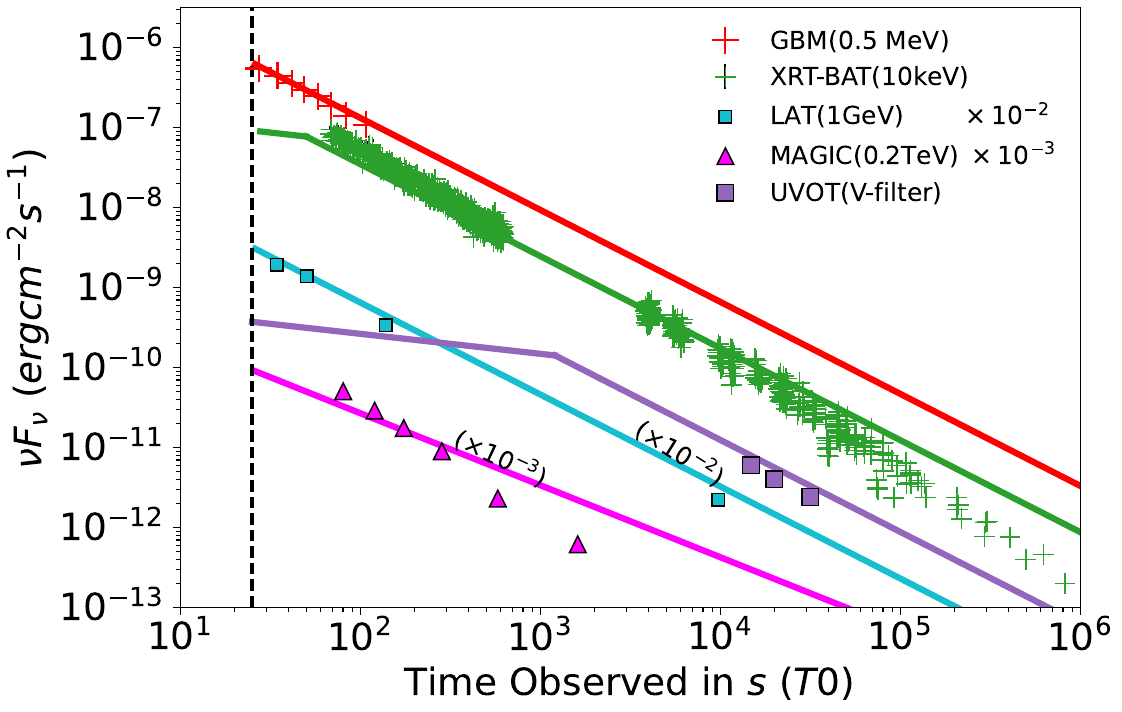}\caption{{Afterglow light-curve in several energy bands, from X-ray to TeV.  The parameters for case (iii) are used to compute the model. { These are $E_{53} = 30$, $n_0 = 15$, $\epsilon_{p,-1} = 8$, resulting in $ \alpha = 41.45$, $\epsilon_{B,-2} = 12$ and $\epsilon_{e,-1} = 0.047$.} The model and data for the LAT and MAGIC band are shifted by factor $10^{-2}$ and $10^{-3}$ respectively. The dashed line shows the division between the prompt phase ($T_0 \leq 25$~s) and the afterglow ($T_0 > 25$~s).}  }
    \label{fig:lightcurve}
\end{figure}

In addition to the data in the two time bins [$\rm T_0 + 68- T_0 + 110$] and [$\rm T_0 +110- T_0 + 180$], MAGIC data exists for three other time bins, namely
$180-360$~s, $360-625$~s and $625-2400$~s \cite{MAGIC:2019irs}. As a consistency
check of our model, we examined the evolution of the proton-synchrotron component
with time, using the same physical parameters, and the self-similar solution to
obtain the dynamical evolution of the Lorentz factor, energy density and magnetic
field, from which the evolution of the flux and characteristic frequencies are
readily obtained. The resulting synchrotron emission from the proton component at
later times is presented in Figure \ref{fig:three graphs} alongside the MAGIC data. {Further, the multi-band light curve of the model in comparison with the observed data of GRB 190114C is shown in Figure \ref{fig:lightcurve}. Here, the emission in the X-ray and GeV bands are obtained for the electron-synchrotron process (which operates in the fast cooling regime) while the emission in the MAGIC band is explained by the proton-synchrotron process (in the slow cooling regime).  }
The similarity between the model results and the data is another independent indication for the ability of our hybrid model to provide acceptable explanations to the data at different times.


\section{Additional radiative processes}
\label{sec:5}
\subsection{Inverse Compton Scattering}

The high energy emission of several GRBs  have been interpreted as originating from the synchrotron
self-Compton process \citep{Dermer_2000, 2001ApJ...548..787S,
2009ApJ...703..675N, liu2013interpretation, 10.1093/mnras/stw1175,
fraija2019synchrotron, Derishev:2021ivd}. This process considers the
inverse Compton (hereinafter IC) interaction of synchrotron photons with the electrons that emitted them. Indeed, it is possible to interpret the observed TeV data using {electron-IC} rather than proton synchrotron, if one assumes different values of the free model parameters, as follows.

The break frequencies for the {electron-IC} component can be obtained within the Thompson
regime, since Klein-Nishina corrections are important only when $\gamma_e h\nu \gtrsim m_e c^2$, see \citet{2001ApJ...548..787S}. The $\nu F_{\nu}$ spectral
peak is achieved at frequency $\nu_{IC,\min}= 2 \gamma_{e,\min}^2 \nu_{e,\min} $ or
$\nu_{IC,c}= 2 \gamma_{e,c}^2 \nu_{e,c}$ in the slow and fast cooling
regimes respectively. For the asserted model specifications (\ref{sec:sync}), these frequencies are defined as
\begin{align}
    h\nu_{IC,\min} &= {7.24 \times 10^{5}\hspace{0.3em}f(p)^2 E_{53}^{3/4} n_{0}^{-1/4} t_{day}^{-9/4} \epsilon_{B,-2}^{1/2} \xi_e^{-4} \epsilon_{e,-1}^4 \text{~eV}}, 
\end{align}
and
\begin{align}    
    h\nu_{IC,c} &= {2.28 \times 10^{8 }\hspace{0.3em}E_{53}^{-5/4} n_{0}^{-9/4} t_{day}^{-1/4} \epsilon_{B,-2}^{-7/2} \text{~eV}} .
\end{align}
The peak-flux of the {electron-IC} spectrum is estimated as \citep{2001ApJ...548..787S} 
\begin{align}
F_{IC,\nu_{\rm peak}} = \frac{1}{3}\sigma_T n_e r F_{e, \nu_{\rm peak}} \approx {1.5 \times 10^{-5}\hspace{0.3em} \xi_e^2 E_{53}^{5/4} n_0^{5/4} t_{day}^{1/4} D_{28}^{-2} \epsilon_{B,-2}^{1/2} \text{~mJy}}.
\end{align}
\begin{figure}[t!]
    \centering\includegraphics[width=0.5 \textwidth]{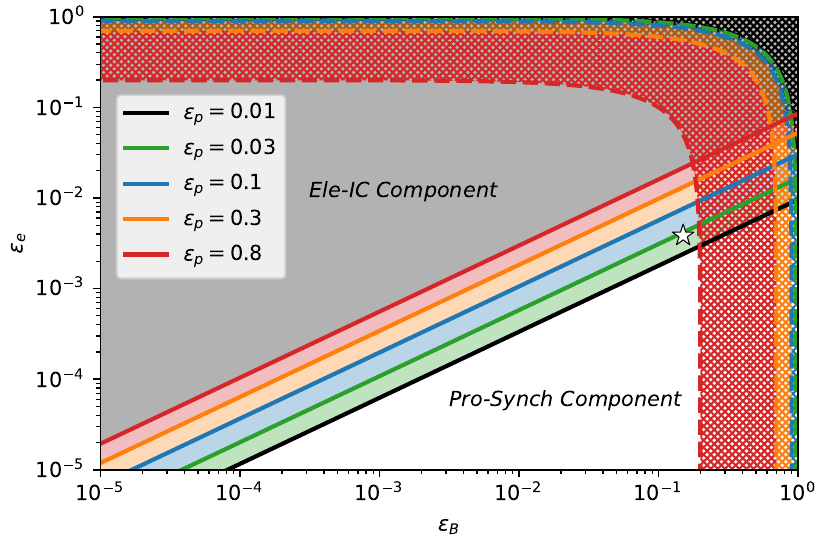}
    \caption{
    {Parameter space for the dominant component at 0.2 TeV. In the grey region, {electron-inverse Compton dominates where $\epsilon_B\ll \epsilon_e$}, {while proton synchrotron dominates  below the solid lines if $\epsilon_B \gg \epsilon_e$}, which colors correspond to the value of $\epsilon_p$ as indicated in the legend. 
    The other parameters are set to $p = 2.2, \xi_e = \xi_p = 0.026, E_{53} = 40$ and $n_0 = 80$ (case (i)). 
    The forbidden zones for each value of $\epsilon_p$ is shown with the hatched regions in their respective colors.  It is apparent that a proton-synchrotron model requires small $\epsilon_e$ and large $\epsilon_p$. The star represents the parameters for case (i). }}
    \label{fig:PS_vs_IC}
\end{figure}
   We can now define the condition for which the proton synchrotron component overcomes the {electron-IC component} (at~$t = 90$~s)  as follows
\begin{align}
    {\frac{ F_{p,\nu}}{ F_{IC,\nu}} = 1.64 \times 10^{-6}E_{53}^{0.225}\epsilon_{B,-2}^{1.75} \epsilon_{p,-1}^{1.2}n_0^{0.525} \xi_e^{0.4}\epsilon_{e,-1}^{-2.4}\xi_p^{-0.2}> 1.} 
\label{eq:PS_vs_IC}
\end{align}
In Figure \ref{fig:PS_vs_IC}, we show the regions in which proton synchrotron or {electron-inverse Compton} dominate the spectrum as a function of $\epsilon_e$ and $\epsilon_B$. The conditions given by Equation \ref{eq:PS_vs_IC} is shown by the solid lines.

{ Equating the electron-IC flux to the MAGIC flux at 0.23 TeV
gives the following }constraints on the parameters of this model as
\begin{align}
\xi_e^2 E_{53}^{\frac{5}{8}} n_0^{\frac{1}{8}} \epsilon_{B,-2}^{-\frac{5}{4}} = 0.236,
\end{align}
{ from which we get}
\begin{align}
\epsilon_{B,-2} = 3.17~E_{53}^{\frac{1}{2}} n_0^{\frac{1}{10}} \xi_e^{\frac{8}{5}}. 
\end{align}
{ This shows that an electron-IC model for explaining the MAGIC observations requires small $\epsilon_B$, in contrast to a proton synchrotron model.}

Also as depicted in figure \ref{fig:PS_vs_IC}, it is evident that the proton-synchrotron process gains significance when $\epsilon_p$ is large. 

However, within the framework of the proton-synchrotron model, for the parameters we considered to explain the data of GRB 190114C in the section \ref{best fit}, the energy peak
flux of the electron-IC spectrum is around $\sim 40$~TeV (90 s) and $\sim 21$~TeV (120 s)  with
the specific flux of $\sim 8.3 \times 10^{-10}$~ergs~cm$^{-2}$~s (90~s) and  $\sim6.2\times 10^{-10}$~ergs~cm$^{-2}$~s (120~s), respectively. The expected contribution from the electron-inverse Compton scattering process is shown
in Figures \ref{fig:xie = xip graphs}, \ref{fig:xie neq xip graphs} and \ref{fig:two graphs}, corresponding to the three models we presented. These figures
show that the {electron-IC component} is always subdominant with respect to the proton
synchrotron component in the TeV band. This was expected since our model requires $\epsilon_{B} \gg \epsilon_e$, resulting in a strong suppression of the SSC component. Thus it states that our model presents an alternative
to the electron-IC component models, as suggested by \cite{Derishev:2021ivd}.

\subsection{Photo-pion Production}

Another mechanism to produce photons
with energy around 1 TeV is photon-pion interactions and their subsequent
electromagnetic pair cascade. Pions ($\pi^0\hspace{0.3em}\&
\hspace{0.3em}\pi^+$) originated from the collisions of the high energy
protons and the seed synchrotron photon-field emitted from the electrons,
produce $\gamma-$ray photons accompanied by neutrinos,
$p^+ + \gamma \rightarrow \pi^0 + p^+$ and $p^+ + \gamma \rightarrow
\pi^+ + n$. Neutral pions quickly decay to photons, $\pi^0 \rightarrow
\gamma + \gamma$, and charged pions decay to positrons
and neutrinos in the decay chain ($\pi^+ \rightarrow \mu^+ + \nu_\mu$, $\mu^+
\rightarrow e^+ +\nu_e + \Bar{\nu}_\mu$). High energy electrons and 
positrons ($e^+$) generated from the muon ($\pi^\pm$) decay
as well as muons and charged pions can emit a
substantial amount of synchrotron radiation. We seek here to compute the
relevance of this process.

In appendix \ref{sec:appendix_ppi}, we derive the expression for the cooling time
by photohadronic interaction, given by Equation \eqref{eq:t_ppi}. This time is
compared to the cooling time by synchrotron given by Equation
\eqref{eq:synch_cooling}:
\begin{align}
    {\left. \frac{t_{synch}}{t_{p \pi}} \right |_{\gamma_{p,\max}} = 2.9 \times 10^{-8} \left( \frac{p-2}{p-1} \right ) E_{53}^{45/32} \epsilon_p^{15/8} n_0^{37/32} t_{\rm day}^{9/8} \xi_p^{-5/16}}
\end{align}
were we also use the constraints on $\epsilon_B$, $\epsilon_e$ and $\xi_e$ given by Equations \eqref{eq:epsilonB}, \eqref{eq:xi_e} and \eqref{eq:epsilone}.
Therefore, for the protons, the synchrotron cooling rate is orders of magnitude faster than the photohadronic interaction, and most
of the energy radiated by the protons is done via synchrotron radiation. {Direct production of photons at 1TeV by $\pi_0$ decay is also subdominant since the energy deposition rate is dominated by the highest proton energy for our choice of proton power law index.}  In other words, 
the contribution from the hadronic cascade in our scenario can be neglected and no modification
of the TeV component is expected. This also means that, for our model, the neutrinos fluence at PeV energy is expected to be small, challenging observations and constraints by IceCube and future instruments.

\section{Discussion} \label{sec:discussion}
\subsection{Radiative efficiency}

As we show here, our proton synchrotron model is able to explain the VHE emission of GRB 190114C. Within the framework of this model, we find that the prompt phase signal has an energy conversion efficiency of a few percent. Indeed,
our most favorable model with $\xi_e = \xi_p$ and $p_e \neq p_p$ requires
a blast-wave with energy $3\times 10^{54}$ ~ergs, resulting in a prompt efficiency
around eight percents. The other models we considered all have a few percent efficiency as well.
Interestingly, this efficiency is comparable to the radiative efficiency predicted in the
internal shock scenario used by many authors to explain the prompt phase; see e.g. \cite{KPS97, PSM99, GSW01}. 

At first, our model requirements on the efficiency seem inconsistent with 
previous findings, specifically for the burst observed by Fermi LAT for which a high
(around 50\%) prompt radiative efficiency is usually determined, \textit{e.g.}
\cite{CFH11}. This trend seems to also be retrieved for bursts observed by the
Neils Gherel Observatory \citep{CFH10}, as well as older bursts \citep{YHS03}. However, we note that:
\begin{itemize}
    \item [(i)] the analysis presented in those papers heavily relies on the
numerical coefficients chosen for the emission  process. Using updated coefficients,
\cite{FP06} re-evaluated the efficiency of several bursts, finding it to be lower,
in the order of few percents, so of similar magnitude than the requirements from our model. 
    \item [(ii)] their analysis also further relies on assuming that all the electrons participate in the radiative process, namely $\xi_e =1$, resulting in an
under-evaluation of the kinetic energy of the blast-wave compared to the case $\xi_e < 1$ as seen from rearranging Equation \eqref{eq:xi_e},
\begin{equation}
    E_{53} \simeq 7.38 \quad \xi_p^{\frac{2}{37}} \epsilon_{p,-1}^{-\frac{12}{37}} n_0^{-\frac{5}{37}} \xi_e^{-\frac{32}{37}},
\end{equation}
for the parameters that satisfy the relation \eqref{eq:epsilonB} at $t = 90$~s.
Note that our proton synchrotron model forbids $\xi_e = 1$, as it would necessarily results in $\epsilon_B > 1$. Instead, our model requires $\xi_e$ to be in the order of $10^{-2}$.
Recent studies analysed the spectral effect of a thermal population resulting in
an incomplete acceleration of particles, see \textit{e.g.} \citet{WBI18, WDB22}. 
Yet, it remains to understand how the modification to the spectral energy distribution impacts the recovery of the blast-wave parameters. Thus, one can conclude that currently there is still no reliable measurement of a high efficiency during GRB prompt emission.
\end{itemize}

\subsection{Constraints on the electron injection fraction.}

The strong constraint on the magnetic field equipartition parameter $\epsilon_B$, requires that the number of electrons participating in the radiation process is smaller than unity, of the order of $10^{-2}$, weakly sensitive to all parameters but the kinetic energy of the blast-wave, $E_{53}$. The 5.5 keV break and the low-energy slope below it, suggest that the thermal component made of the bulk of the electrons, should have a temperature much smaller than $\gamma_{\min}$. If this was not the case, the low-energy slope below $\nu_{\min}$ would be different and entailed to the exact injection function.

In other words, we require an acceleration scenario in which the thermal component and accelerated particles are two clearly separate entities. Such an injection
function was originally proposed by \citet{2005ApJ...627..861E}. In that paper, they
also studied the effect of this assumption on the emission properties of GRB
afterglow. In particular they found that such model can be constrained by early
afterglow emission, underlying the necessity of multi-wavelength observations. 
Multi-wavelength observations of GRB 190114C afterglow, which include the temporal evolution in the radio and optical bands, was reported by \citet{2021MNRAS.504.5685M}. And indeed, the constraint upon the parameter $\epsilon_e$ at time interval of 65~s reported in that work, resulted with an electron injection fraction of 2$\%$, which is comparable to the fraction $\xi_e$ obtained in our model. These consistent results therefore serve as an independent support to our model.

{ The presence of a dominant (in number) thermal electron population would leave a visible footprint in the spectrum at low energies. In our analysis, since we require $\nu_m < \nu_{XRT}$, such footprint could be found in the optical and radio bands. The emission of those thermal electrons is usually neglected \cite[e.g.][]{2021MNRAS.504.5685M}. Such hypothesis could in principle be tested and its parameters constrained by early X-ray and MeV observations in long GRBs with duration $>$ 100~s. To the best of our knowledge this has not been done yet.}

\subsection{Constrain on the jet energetic from the jet opening angle}
The X-ray light-curve of GRB 190114C is well described with a power-law decay of the flux until at least $10^6$~s~\footnote{ The XRT light-curve repository reports a change of slope around $t = 5 \times 10^4$. Using this time as the jet break would results in lowering the energy inferred below by about 1 order of magnitude.}. From the derived model parameters, we can therefore infer a lower limit on the jet opening angle and subsequently on the jet energy.
The opening angle ($\theta$) of the jet depends upon three quantities, which are the isotropic-equivalent 
kinetic energy, $E_{k}$, the jet-break time, $t_{break}$ and the ambient 
number density,  $n$ \citep{2005ApJ...629L..13L}. From 
our model (case iii), we derived $E_k = 30 \times 10^{53}$~erg and $n = 15$~cm$^{-3}$. Hence, a lower limit on $\theta$ is given by 
\begin{equation}
\theta > {7.2 \times 10^{3} \left(\frac{t_{break}}{(1+z)}\right)^{3/8}\left(\frac{ n}{E_{k}}\right)^{1/8} \approx 0.24~\rm rad.}
\end{equation} 
Therefore, the collimation-corrected energy of the jet is 
\begin{equation}
    E_{jet} > \frac{\theta^2}{2}E_{k} \approx 8.85 \times 10^{52}~\rm erg.
\end{equation}
For case (ii), when  $E_k = 50 \times 10^{53}~\rm erg$ and $n = 130$~cm$^{-3}$ we get $\theta > 0.3 $~rad and $E_{jet} \gtrsim 2.23 \times 10^{53} \rm ~erg$. In the same way, for case (i) we derived $E_k = 40 \times 10^{53}~\rm erg$ and $ n = 80$~cm$^{-3}$, hence $\theta > 0.3$~rad and $E_{jet} \gtrsim 1.67 \times 10^{53}~\rm erg$.

{ The obtained jet energy appears to be large when comparing to other estimates  which finds a typical energy output in the order of $10^{51}$~ergs \citep{BND15, WZL15, WZL18}. 
We note however that these estimates rely on afterglow modeling with the assumption $\xi_e = 1$, while the total energy of the burst scales $\propto \xi_e ^{-1}$.}
On the other hand, 
{relativistic jets from solar mass black holes (microquasars) transfer a significant amount of their kinetic energy, estimated to be $\sim 10^{51}$ erg, to the surrounding ambient medium \citep{1998AJ....116.1842D, 2003IAUS..214..201M}. Our model therefore requires a jet energy to be larger by a factor of $\approx$ tens to hundreds than the energy released by microquasar jets.} 

\subsection{Production of high-energy protons}

{
To explain the MAGIC observations with proton synchrotron, protons have to be accelerated to comoving energies around $10^{19}$~eV, which results in observed particle energies around $10^{21}$~eV. This requirement would make GRBs able to accelerate the highest energy ultra-high energy cosmic rays (UHECR) observed on Earth (though most of them would not be detected due to the GZK cutoff). It is well known that GRBs satisfy the Hillas criterion \citep{Hil84} making them a plausible source of UHECRs.  We further note that the acceleration mechanisms must results in proton power-law distribution function formed over many orders of magnitude in energies. 

If such a challenging acceleration can indeed take place or not is not certain. \citet{SSA13} performed PIC simulations of particle acceleration in collisionless plasma and estimated the maximum proton Lorentz factor for an external shock in the context of GRBs. They found that this maximum Lorentz factor is $\gamma_{\rm max} \approx 10^8$, weakly dependent on the parameters. This is three orders of magnitude lower than the proton Lorentz factor required by our parameters. This result is however based on the scaling law $\gamma_{\rm max} \propto t^{1/2}$. More recent results tentatively obtained the scaling law $\gamma_{\rm max} \propto t$ \citep{HRK23}. For this temporal scaling, the production of the highest energy protons required by our model is possible.
}
\section{Conclusion}

\label{sec:Conclusion}

The broken power-law spectrum of the electron-synchrotron emission  model is a
standard prediction of GRB afterglow theories, which was successful in explaining the spectra at energy lower than a few MeV. Yet, the source of the VHE afterglow component is still being
investigated. GRB 190114C is one such a burst with a high energy ($0.2 \sim 1$~TeV) peaking
afterglow spectral component as reported by MAGIC within the epoch $60$~s to $2400$~s. We argue here that the source of the VHE segment
is of proton synchrotron origin, while the sub-MeV component is explained by
electron synchrotron radiation within the framework of the classical fireball
evolution scenario. {According to the model discussed in this paper, protons are simultaneously accelerated with the electrons in the blast wave, allowing protons to radiate in the (sub-)TeV band. }

We provided the constraints that this model must satisfy under different 
conditions for the parameters describing the uncertainty of the particle acceleration process. We presented three models
with different injection parameters characterizing the particle acceleration and injection : (i)
our first model has similar fractions of electrons and protons accelerated to a high-energy power law, $\xi_e = \xi_p = 2.6 \times 10^{-2}$; (ii) the second one has $\xi_e \neq \xi_p $ and (iii) the last one has $\xi_e = \xi_p = 2.3\times 10^{-2}$, but (slightly) different power law indices, $p_e = 2.2$ and $p_p = 2.1$, that is to say electrons and protons are described by a
different injection index. All the scenarios in our model exhibit external medium density $n \sim 10 - 100$~cm$^{-3}$, which is somewhat higher than the fiducial value often assumed in explaining many GRB afterglows, $n \approx 1$~cm$^{-3}$, but not unreasonably high. In fact a similar value was inferred from fitting the much later time radio and optical afterglow data  \citep{2021MNRAS.504.5685M}.  Similarly, a reasonably high blast-wave
kinetic energy ($E \approx 3\times10^{54}$~erg) which, given the inferred energy from the observed photons of $\approx 2.5 \times 10^{53}$~erg during the prompt phase, leads to a considerable prompt
radiative efficiency of $\approx 8\%$.  We point out that this value of efficiency is similar to the efficiency expected in kinetic energy conversion by internal shocks. Finally, we note the degeneracy in determining the values of the explosion energy and the ambient density: for a given observed flux, a low ambient density can be compensated by a higher explosion energy for a fixed magnetization, $\epsilon_B$, thereby reducing the radiative efficiency of the prompt phase.

\citet{2021MNRAS.504.5685M} analysed and interpreted the radio, optical and X-ray data
of GRB 190114C at both early and later times. Very interestingly, they concluded that the kinetic energy $E$
was at least one order of magnitude higher than the observed isotropic
equivalent energy. This result is in excellent agreement with
the constraints on the energy we obtained here. Further examination of
our model requirements conveyed low { requirements on the
normalisation of the} acceleration rate with $\alpha > 10$.
In addition, we find that in order to reproduced the TeV band data via the
proton synchrotron process, the fractions of both electrons and protons
accelerated to a high energy power law, $\xi_e$ and $\xi_p$ are at the
order of few percent. It also requires a large magnetic field, hence a
large kinetic energy, from which, using the observed data at the
{\it Swift-XRT} band, a constraint on the fraction of electrons
accelerated, $\xi_e$ is imposed as well. 

 The injection fractions we derive, of a few percents, are similar to the ones derived using the radio, optical and X-ray observational data in \citet{2021MNRAS.504.5685M}. Furthermore, these values are consistent with 
the theoretical values found in particle-in-cell simulations of particle acceleration in relativistic shock waves \citep{2008ApJ...682L...5S}. Using the same values, we showed in Figure \ref{fig:three graphs} that the proton-synchrotron mechanism is in agreement with the MAGIC data at later times as well, until the last existing MAGIC observation at 2400~s.  Importantly, we showed that for the constraints set by the data { on the model presented here}, the proton synchrotron emission is dominant at the TeV band over other emission processes such as SSC and photo-pion production mechanism. This makes this model a viable alternative to leptonic models.

{ Our model also suffers from a certain number of limitations. First, it is
not clear if protons can be accelerated to the energies
($\sim 10^{20}$~eV in the observer frame) which are required
to explain the TeV emission. Second, the beaming corrected energy deduced
for this burst is one to two orders of magnitude larger than the energy
inferred for GRBs, since protons energy density and magnetic field energy
density are required to be large. We however point out that those estimates
are model dependent and rely on the assumption $\xi_e = 1$. Finally, we
discarded the radiation from the thermal electron, which should emit a synchrotron component around the optical band. SSC models do not suffer from these drawbacks but instead are challenged by the broadband modeling specifically in the optical \citep[for a thorough summary, see section 4.3 of][]{MN22}.  }

To conclude, we presented a proton synchrotron model to
explain the VHE afterglow spectrum of GRB 190114C and found that it can accommodate the
available observed MAGIC data-set from the epoch of 68 s up until 2400~s. The parameters we found are consistent with both independent measurements, as well as theoretical predictions of particle acceleration. Furthermore, the uncertain values of the explosion energy and ambient density, are within one-two orders of magnitude from 'fiducial' values often assumed in the literature. These results, therefore, point to the need of a more thorough investigation of the role of proton-synchrotron emission during the afterglow phase in GRBs.

The VHE band limit of the GRB afterglows is anticipated to
reach higher magnitudes with CTA in the near future. Therefore, highlighting the role
played by highly relativistic protons in the GRB afterglow theory is of crucial
importance. Conversely, the acceleration mechanism and radiative efficiency of protons are expected to
be further constrained by the observation of very-high energy emission ($\sim 1$~TeV)
in the afterglow spectrum.

\bibliography{sample631}{}
\bibliographystyle{aasjournal}

\appendix

\section{Contribution of photohadronic emission}

\label{sec:appendix_ppi}
In this appendix, we compute the cooling rate by photohadronic interaction for parameters
relevant for the proton synchrotron model presented herein. Since at frequencies larger
than $\nu_{\min}$, the photon numbers from the electron
synchrotron component falls off rapidly, we neglect their contribution to the
photohadronic interaction. Therefore, we compute the Lorentz factor $\gamma_p^{th}$ of
the protons interacting with electron synchrotrons at the comoving frequency of the peak flux,
$\nu_{\min}^{'} = \nu_{\min} / \Gamma$, where $\nu_{\min} = 5.5$~keV, by using the threshold condition of photohadronic
interactions $2 \gamma_p \nu_{\min}^{'} = 132$ MeV :
\begin{align}
    \gamma_p^{\rm th} = 1.2\times 10^5 ~ E_{53}^{\frac{1}{8}} n_0^{-\frac{1}{8}} t^{-\frac{3}{8}}_{day}
\end{align}
Comparing the threshold Lorentz factor to the maximum proton Lorentz factor given in 
Equation \eqref{eq:proton_gamma_max}, it is clear that $\gamma_p^{th} < \gamma_{p,\max}$,
meaning that the protons at $\gamma_{p, \max}$ producing the TeV component can also interact
with low-energy photons produced by electron synchrotron. Therefore, we now estimate the
efficiency of the photohadronic process.

The proton cooling rate for the photo-pion production is given by \citet{MK94}, see
also \citet{BRS90,WB97} and reads 
\begin{equation}
    t_{p\pi}^{-1} = \frac{c}{2\gamma_p^2} \int_{\Bar{\epsilon}_{th}}^{\infty} d\Bar{\epsilon} K_p(\bar \epsilon) \sigma_{p\pi}(\bar \epsilon) \Bar{\epsilon} \int_{\frac{\Bar{\epsilon}}{2\gamma_p}} dx \frac{n_x(x)}{x^2} \label{eq:cooling_photo_pion_general_eq}
\end{equation}
Here $x$ is the photon energy in units of the electron rest mass energy and
$n_x$ is the comoving photon spectral number density produced by the electron synchrotron
process. Since electrons are fast cooling in our model (the magnetic field needs to be large),
we obtain $n_x$ as \citep{Sari_1998}
\begin{align}
    n_x = \frac{u_{\nu_c}}{h x} \left \{
    \begin{aligned}
    &\left (\frac{x}{x_c} \right )^{\frac{1}{3}}  & & x < x_c \\
    &\left (\frac{x}{x_c} \right )^{-\frac{1}{2}}   &  &  x_c < x < x_m \\
    &\left (\frac{x_m}{x_c} \right )^{-\frac{1}{2}} \left ( \frac{x}{x_m} \right )^{-\frac{p}{2}}   &  ~~~~ &    x_m < x 
    \end{aligned} \right.
\end{align}
where 
\begin{align}
    u_{\nu_c} \sim \frac{n_e P_{\rm synch} (\gamma_{e,c})}{\nu_{e,c}} t_{\rm dyn} = 1.86 \times 10^{-15}~ E_{53}^{\frac{3}{8}}\epsilon_{B, -2}^{\frac{1}{2}} n_0^{\frac{9}{8}} t_{day}^{-\frac{1}{8}} {\rm ~erg~cm^{-3}~Hz^{-1}}
\end{align}
To simplify analytically the expression given in Equation
\eqref{eq:cooling_photo_pion_general_eq}, we follow \cite{petropoulou2015bethe} and set :
\begin{align}
    \sigma_{p\pi} &= \sigma_0 H(\bar \epsilon - \bar \epsilon_{\rm th}) \\
    \sigma_0 &= 1.5 \times 10^{-4} \sigma_T \\
    K_p &= 0.2 \\
    \bar \epsilon_{\rm th} &= 145 {\rm~ MeV,}
\end{align}
where $H$ is the Heaviside function. We simplify this expression for protons with 
Lorentz factor $\gamma_{p, \max}$, which are the ones responsible for producing the 
spectral component in the TeV band. Moreover, for most of the parameter space, we have
\begin{align}
2 \gamma_{p,\max} \nu_{e,c}^{'} =  2.3 \times 10^{-8}  E_{53}^{\frac{67}{32}} \epsilon_p^{\frac{21}{8}} n_0^{\frac{3}{32}} t_{\rm day}^{\frac{1}{16}} \xi_p^{-\frac{7}{16}} {\rm ~MeV~} \ll 132 {\rm ~MeV,}
\end{align}
where we used Equations \eqref{eq:epsilonB}, \eqref{eq:xi_e} and \eqref{eq:epsilone}. In this equation, $\nu_{e,c}^{'} = \nu_{e,c} / \Gamma$ is the comoving cooling frequency.
Therefore the main contribution to the integral in Equation
\eqref{eq:cooling_photo_pion_general_eq} is for photons with frequency
between $\bar \epsilon_{th}$ $\text{~and~} \nu_m$. We now turn to the computation of the integral :
\begin{align}
    \frac{2\gamma_p^2}{c K_p \sigma_0} t_{p\pi}^{-1} &= I_2 + I_3
\end{align}
where 
\begin{align}
   I_2 &= \int_{\bar \epsilon_{\rm th}}^{2 \gamma_p x_m} d\Bar{\epsilon}  \Bar{\epsilon} \int_{\frac{\Bar{\epsilon}}{2\gamma_p}}^\infty dx \frac{n(x)}{x^2} \\
   I_3 &= \int_{2 \gamma_p x_m}^\infty d\Bar{\epsilon}  \Bar{\epsilon}\int_{\frac{\Bar{\epsilon}}{2\gamma_p}}^\infty dx \frac{n(x)}{x^2}
\end{align}
In other word the integral has 2 contributions for different values of $\bar \epsilon$. We have
\begin{align}
    I_2 = \frac{u_{\nu_c}}{h} (2 \gamma_p)^2 \left ( \frac{x_c}{x_m} \right)^{\frac{1}{2}} \left [  \frac{2p-2}{10+5p} \left( \left (\frac{\bar \epsilon_{\rm th}}{2\gamma_p x_m} \right )^2 - 1\right)  + \frac{4}{5} \left( \sqrt{ \frac{2\gamma_p x_m}{\bar \epsilon_{\rm th}} } - 1 \right )  \right ] 
\end{align}
which in the limit $\bar \epsilon_{\rm th} \ll 2 \gamma_p x_m$ reduces to
\begin{align}
    I_2 \sim \frac{4}{5} \frac{u_{\nu_c}}{h}  (2\gamma_p)^2 \sqrt{\frac{x_c}{x_m}} \sqrt{\frac{2 \gamma_p x_m}{\bar \epsilon_{\rm th}}}.
\end{align}
Finally for $I_3$, we find
\begin{align}
    I_3 = \frac{u_{\nu_c}}{h}  \sqrt{\frac{x_c}{x_m}} \frac{1}{4 + p} (2\gamma_p^2)
\end{align}
Therefore it is cleat that $I_2$ dominates the contribution to the integral. Numerically, for $\gamma_{p,\max}$, we obtain for the cooling time by photohadronic interaction 
\begin{align}
        t_{p\pi}^{-1} \approx 1.2 \times 10^{-8}  \left ( \frac{p-2}{p-1} \right ) \epsilon_{e,-1} n_0^{\frac{15}{16}}  t_{\rm day}^{\frac{9}{16}} E_{53}^{-\frac{3}{16}} \alpha^{-\frac{1}{2}}  \epsilon_{B,-2}^{-\frac{1}{4}} \xi_e^{-1} {\rm ~s^{-1}~} \label{eq:t_ppi}
\end{align}

\end{document}